\documentclass[
    ,conference
]{IEEEtran}
\IEEEoverridecommandlockouts
\usepackage{amsmath, amsthm}
    
\usepackage{graphicx}
\usepackage[table]{xcolor} % xcolor instead of color
\usepackage{hyperref, url}
\usepackage{adjustbox}
\usepackage{booktabs, multirow}
\usepackage{caption}
\usepackage{subcaption}
\usepackage{enumitem}
\usepackage{lipsum}
\usepackage{setspace}
\usepackage[normalem]{ulem}

\usepackage{algorithm, algpseudocode}
    \definecolor{BLUE}{rgb}{.0, .2, .6}
    \definecolor{BLUEalt}{HTML}{1e50a2}
    \definecolor{RED}{HTML}{c9171e}
    \algrenewcommand{\alglinenumber}[1]{{\scriptsize\bfseries\ttfamily\color{RED}#1}}

\usepackage{thmtools}
\declaretheoremstyle[%
  spaceabove=4pt,%
  spacebelow=4pt,%
  headfont=\normalfont\itshape,%
  postheadspace=1em,%
  qed=\qedsymbol%
]{mystyle} 
\declaretheorem[name={Proof},style=mystyle,unnumbered,
]{prf}

\usepackage{footnote}
\makesavenoteenv{tabular}
\makesavenoteenv{table}

\usepackage{xspace}

\usepackage{tikz}
\usepackage{array}
\usepackage{booktabs}
\usepackage{tabu}

\usepackage[zerostyle=d]{newtxtt}

\usepackage{cite}
\usepackage{amsmath,amssymb,amsfonts}
\usepackage{graphicx}
\usepackage{textcomp}
\usepackage{comment}
\usepackage{xcolor}
\usepackage{threeparttable}
\def\BibTeX{{\rm B\kern-.05em{\sc i\kern-.025em b}\kern-.08em
    T\kern-.1667em\lower.7ex\hbox{E}\kern-.125emX}}
\begin{document}

\title{Improving Prediction-Based Lossy Compression Dramatically via Ratio-Quality Modeling}

\newcommand{\BetterMark}{$^\star$}

\author{
Sian Jin\BetterMark,
Sheng Di\IEEEauthorrefmark{2},
Jiannan Tian\BetterMark,
Suren Byna\IEEEauthorrefmark{3},
Dingwen Tao\BetterMark\thanks{Corresponding author: Dingwen Tao (\url{dingwen.tao@wsu.edu}), School of EECS, Washington State University, Pullman, WA 99164, USA.},
Franck Cappello\IEEEauthorrefmark{2}\\
\BetterMark%
School of Electrical Engineering and Computer Science, Washington State University, Pullman, WA, USA\\
%%%% WSU
\IEEEauthorrefmark{2}%
Mathematics and Computer Science Division, Argonne National Laboratory, Lemont, IL, USA\\
%%%% ANL
\IEEEauthorrefmark{3}%
Computational Research Division, Lawrence Berkeley National Laboratory, Berkeley, CA, USA%\\
%%%% UIUC
%\IEEEauthorrefmark{5}University of Illinois at Urbana-Champaign, IL, USA
}

\maketitle

\begin{abstract}

Error-bounded lossy compression is one of the most effective techniques for reducing scientific data sizes.
However, the traditional trial-and-error approach used to configure lossy compressors for finding the optimal trade-off between reconstructed data quality and compression ratio is prohibitively expensive. To resolve this issue, we develop a general-purpose analytical ratio-quality model based on the prediction-based lossy compression framework, which can effectively foresee the reduced data quality and compression ratio, as well as the impact of lossy compressed data on post-hoc analysis quality.
Our analytical model significantly improves the prediction-based lossy compression in three use-cases:
% \textcolor{black}{(1) data management optimization for scientific applications}
(1) optimization of predictor by selecting the best-fit predictor; 
(2) memory compression with a target ratio; 
and (3) in-situ compression optimization by fine-grained tuning error-bounds for various data partitions.
We evaluate our analytical model on 10 scientific datasets, demonstrating its high accuracy (93.47\% accuracy on average) and low computational cost (up to 18.7$\times$ lower than the trial-and-error approach) for estimating the compression ratio and the impact of lossy compression on post-hoc analysis quality. 
We also verify the high efficiency of our ratio-quality model using different applications across the three use-cases.
\textcolor{black}{In addition, our experiment demonstrates that our modeling-based approach reduces the time to store the 3D RTM data with HDF5 by up to 3.4$\times$ with 128 CPU cores over the traditional solution.}
%our model contributes to data management optimization for scientific applications and provides up to 3.4$\times$ time reduction when handling the data.}

\end{abstract}

\maketitle
\pagestyle{plain}
\pagenumbering{gobble}

\setlength{\textfloatsep}{6pt}
\setlength{\abovecaptionskip}{3pt}
\setlength{\abovedisplayskip}{2pt}
\setlength{\belowdisplayskip}{2pt}
\setlength{\abovedisplayshortskip}{2pt}
\setlength{\belowdisplayshortskip}{2pt}

\section{Introduction}

Large-scale scientific simulations on parallel computers play an important role in today's science and engineering domains.
Such simulations can generate extremely large amounts of data. % that are highly compute and storage intensive.
For example, one Nyx~\cite{almgren2013nyx} cosmological simulation with a resolution of $4096\times 4096 \times 4096$ cells can generate up to 2.8 TB of data for a single snapshot; a total of 2.8 PB of disk storage is needed, assuming the simulation runs 5 times with 200 snapshots dumped per simulation.
\textcolor{black}{Despite the ever-increasing computation power can be utilized to run the simulations nowadays, 
%storing and querying 
managing such large amounts of data remains challenging.}
%a major challenge.}
It is impractical to save all the generated raw data to disk due to: (1) limited storage capacity even for large-scale parallel computers, and (2) the I/O bandwidth required to save this data to disk can create bottlenecks in the transmission~\cite{wan2017comprehensive,wan2017analysis,cappello2019use}.

Compression of scientific data has been identified as a major data reduction technique to address this issue.
More specifically, the new generation of error-bounded lossy compression techniques, such as SZ~\cite{tao2017significantly, di2016fast, liangerror} and ZFP~\cite{zfp}, have been widely used in the scientific community ~\cite{di2016fast,tao2017significantly,zfp,liangerror,lu2018understanding,luo2019identifying,tao2018optimizing,cappello2019use,jin2020understanding,grosset2020foresight}. 
Compared to lossless compression that typically achieves only $2\times$ compression ratio~\cite{son2014data} on scientific data, error-bounded lossy compressors provide much higher compression ratios with controllable loss of accuracy.

\textcolor{black}{
%Different from general data management approaches, scientific datasets generated by large-scale simulations are typically managed by using efficient parallel I/O libraries/middleware such as Hierarchical Data Format 5 (HDF5)~\cite{hdf5} for the high-performance purpose.
%format for post-hoc analysis, such as the 
%For example, HDF5 has been considered as the most state-of-the-art scientific data management software for storing and querying scientific datasets \cite{folk2011overview,nyx}. 
Scientific applications on large-scale computer systems such as supercomputers typically use parallel I/O libraries, such as HDF5~\cite{hdf5}, for managing the data. In specific, Hierarchical Data Format 5 (HDF5) is considered to provide high parallel I/O performance, portability of data, and rich API for managing data on these systems. HDF5 has been used heavily at supercomputing facilities for storing, reading, and querying scientific datasets \cite{folk2011overview,nyx}.
This is because HDF5 has specific designs and performance optimizations for popular parallel file systems such as Lustre~\cite{byna2020exahdf5,pokhrel2018parallel}.
%databases for scientific datasets.
%In addition, instead of using general database in distributed storage, these datasets have their specific data management approach based on the parallel file system~\cite{byna2020exahdf5,pokhrel2018parallel}.
Moreover, HDF5 also provides users dynamically loaded filters \cite{hdf5filter} such as lossless and lossy compression \cite{hdf5filter-sz}, which can automatically store and query data in compressed formats.
%thus it allows scientific applications automatically store and query the data in compressed formats, 
HDF5 with lossy compression filters can not only significantly reduce the data size, but also improve 
%Suren (editing): 
%the scientific data management performance.
performance of managing scientific data.
%This allows us to not only significantly reduce the data sizes with scientific lossy compressors, but also reduce the challenges in scientific data management caused by large data size.
}

%In order to efficiently utilize lossy compressors in practice, it is
However, for HDF5 to take advantage of lossy compressors, it is 
essential for users to identify the optimal trade-off between the compression ratio and compressed data quality, which is fairly complex.
Since there is no analytical model available to foresee/estimate the compression quality accurately, the configuration setting (such as error bound types and values) of error-bounded lossy compressors for scientific applications relies on empirical validations/studies based on domain scientists' trial-and-error experiments~\cite{jin2020understanding,grosset2020foresight,tao2018optimizing}. The trial-and-error method~\footnote{The trial-and-error experiment is to compress and decompress the data with different feasible error bounds (or combinations of error bounds) and measure the compression ratio and data quality to choose the best error bound(s).} suffers from two significant drawbacks, which leads to significant issues in practice. First, this method has an extremely high computational cost, in that users need to run applications with diverse combinations of input data, and each run may cost tremendous computational resources. For example, to find an optimized error bound for a given Nyx simulation with a qualified
power spectrum analysis, about 10 trials of compression-decompression-analysis are needed before compressing the data with the optimized configuration~\cite{jin2020understanding,grosset2020foresight}. Second, the identified configuration setting is still dependent on specific conditions and input data, so it cannot be applied across datasets generically because of the lack of a theoretical compression quality model. %As such, the lack of analytical model linking compression ratios and data and post-hoc analysis quality represents a significant gap to address.

%Users often search for the optimal trade-offs by manually tuning compressor parameters, such as error bound types and error bound values. 
%Trials and error method can be long and inefficient. In addition to its high computational cost, this configuration optimization method reduces users' confidence in lossy compressors because it is not linking formally the compression ratio and its impact on the data quality and the post-hoc analysis quality. Moreover, this method presents a challenge for domain scientists who are not familiar with lossy compressors.
% Since error-bound is one of the most important variable for ratio-quality balancing, previous studies use trail-and-error approach to empirically study each lossy compressors' efficiency and to select suitable configuration for a specific application offline~\cite{jin2020understanding,grosset2020foresight,tao2018optimizing}.
%Moreover, HPC computing resources are very important for scientific applications that run at scale, thus from users perspective, 
%the trial-and-error methods to configure the compressors (trying different prediction, quantization, and encoding methods) is competing with the simulation  for the allocated resources.
%Furthermore, while this approach consumes tremendous amount of computational resources, it can only provide close-to-optimal compressor configurations due to the dispersed samples..

In this paper, we theoretically develop a novel, analytical model in terms of the prediction-based lossy compression framework, that can efficiently and accurately estimate the compression quality such as ratio and data distortion for any given dataset, consolidating the confidence of lossy compression quality for users. Specifically, we perform an in-depth analysis for the critical components across multiple stages of the prediction-based error-bounded lossy compression framework, including distribution of prediction errors, Huffman encoding efficiency, effectiveness of quantization, post-hoc analysis quality, and so on. Our model features three critical characteristics: (1) it is a general model suiting most scientific datasets and applications, (2) it has a fairly high accuracy in estimating both ratio and post-hoc analysis quality, and (3) it has very low computational overhead.

%We built our models for each element in the compressor and the post-hoc analysis from a deep understanding of modularized prediction-based lossy compressor.
%This is the first attempt to the best of our knowledge, and it can significantly improve the compression quality for prediction-based lossy compressors.
% we focus on developing theoretical-based models for accurately estimating the compression ratio and post-hoc analysis quality degradation. 

To the best of our knowledge, this work is the first attempt to develop an analytical model theoretically for lossy compression quality. which fundamentally differs from all existing compression modeling approaches.
% \textcolor{black}{Most previous works regarding lossy compression implementation mainly aimed to develop or improve the lossy compression ratio based on a specific algorithm or framework. They mainly focus on how to improve prediction efficiency and/or encoding efficiency~\cite{sz16,sz18,liang2018efficient,sz17,sz18,zhao2021optimizing}. Some other previous works use empirical studies or trial-and-error methods for investigating or improving lossy compression in various scientific applications~\cite{cappello2019use,tao2019optimizing,fraz,sz16} and application-specific optimizations of lossy compression~\cite{jin2020understanding,grosset2020foresight}. In contrast, this paper is the first efficient attempt to provide an accurate estimation of compression ratio/quality and theoretically guide the use of lossy compression in database/scientific applications.
% In addition, our approach is also differs from all existing compression modeling approaches.}
Lu et al. \cite{lu2018understanding}, for example, focus only on the quantization stage and compression ratio estimation by extrapolation, while our model considers all compression stages for both ratio and quality, which can significantly improve the modeling accuracy.
\textcolor{black}{Wang et al. \cite{wang2019compression} developed a simple model to estimate compression ratios based on an empirical study of the correlation between compression ratio and multiple statistical metrics of the data 
%empirical studies and selected metrics, \sout{which is proven to be both computationally intensive and inaccurate for many scientific datasets.} \textcolor{black}{Their approach empirically studied the correlation between compression ratio and multiple statistical metrics of the data 
(such as prediction hit ratio). It can only provide a rough estimation of compression ratio ($\sim$10\%$\sim$60\% error rate), which cannot satisfy the real-world application demand~\cite{grosset2020foresight}.}
Jin et al. \cite{jin2020adaptive} leveraged a simplified error distribution of SZ compressor to estimate the post-hoc analysis quality for cosmology application. The compression quality estimation, however, relies on an empirical study which is specific to Nyx cosmology datasets. Compared to all existing solutions in estimating the lossy compression quality, our proposed model can offer in-situ optimization of lossy compression quality with significantly higher compression ratios and low computational overhead.

% On the one hand, HPC computing resources are very precious to scientific applications that run at scale, thus it is improper to use trial-and-error methods to get to understand the reduction effectiveness---i.e., compression ratio and reconstructed data quality---of different prediction, quantization, and encoding methods. Our proposed ratio-quality modeling can efficiently avoid the expensive checking overheads and enable more fine-grained optimization.
% On the other hand, the usage of error-bounded lossy compression for scientific applications has long relies on empirical study and the experience of scientists~\cite{}, which reduces the confident and enlarge the challenges for scientists who are not familiar with provided lossy compressor on given dataset.
% Our proposed ratio-quality model significantly reduces the empirical learning-curve for given dataset while also provide crucial theoretical support to increase the confidence when deploy lossy compression.
% Different from the existing modeling approach \cite{lu2018understanding} that only focuses on the quantization stage and compression ratio estimation by extrapolation, our modeling method will focus on all stages and both ratio and quality.

%Our proposed ratio-quality model can significantly reduces the empirical learning-curve for a given dataset and also provides crucial theoretical support to increase the confidence of using lossy compression. 

%%% After which, we utilize our proposed model on various scientific data applications to demonstrate the accuracy and effectiveness. 

The contributions of this work are summarized as follows:
\begin{itemize}[noitemsep, topsep=3pt, leftmargin=1.3em]
    % \item We propose a novel modularized ratio-quality model for prediction-based lossy compressors to provide accurate estimates of compression ratio and post analysis quality. In particular, we build a theoretical model to efficiently estimate the encoder efficiency and provide essential parameters for compression ratio estimation. Moreoever, we build a theoretical analysis and a general guideline for post-hoc analysis quality degradation estimation.
    % \item We build a novel modularized ratio-quality model for prediction-based lossy compressors to provide accurate estimates of compression ratio and post analysis quality.
    \item We decouple prediction-based lossy compressors to build a modularized model for ratio and quality estimation. 
    %and modularize  to build the methodology and workflow on modeling the ratio-quality.
    \item We theoretically analyze how to estimate the encoder efficiency and provide essential parameters for compression ratio estimation. We build a fine-tuning mechanism to improve the lossy compression quality estimation accuracy for different predictors.
    \item We propose a theoretical analysis
    %and provide a general guideline 
    to estimate the qualification of lossy decompressed data on post-hoc analysis based on the estimated error distribution considering both uniform and nonuniform distributions.
    \item We evaluate our model using 10 real-world scientific datasets involving 17 fields. Experiments verify that our approach can minimize the overhead of compression optimization and provide accurate ratio and quality estimation.
    \item We evaluate our model on three use-cases and show that it can significantly improve the performance of prediction-based lossy compressors in terms of optimization time overhead and overall compression time for predictor optimization, memory compression optimization, and fine-grained ratio-quality optimization.
    % We demonstrate the effectiveness of our model compared to previous approaches on three optimization  use-cases, including predictor optimization, memory compression optimization, and fine-grained ratio-quality optimization.
\end{itemize}

\begin{comment}
The rest of this paper is organized as follows. In Section~\ref{sec:background}, we discuss the background and challenges. In Section~\ref{sec:analysis}, we describe our proposed ratio-quality model.
%for lossy compression of scientific data with prediction-based lossy compressor. 
In Section~\ref{sec:design}, we present the optimization use-cases of our proposed model. In Section~\ref{sec:evaluation}, we present our evaluation results. % of our proposed model.
% in terms of accuracy and performance, as well as the evaluation on the three optimization use-cases compared with previous parameter optimization approaches. 
In Section~\ref{sec:conclusion}, we conclude our work and discuss our future work.
\end{comment}

\section{Research Background}
\label{sec:background}

In this section, we present the background information on lossy compression and discuss the research challenges.

\subsection{Data Management in Scientific Applications}

\begin{figure}[]
    \centering
    \includegraphics[width=\linewidth]{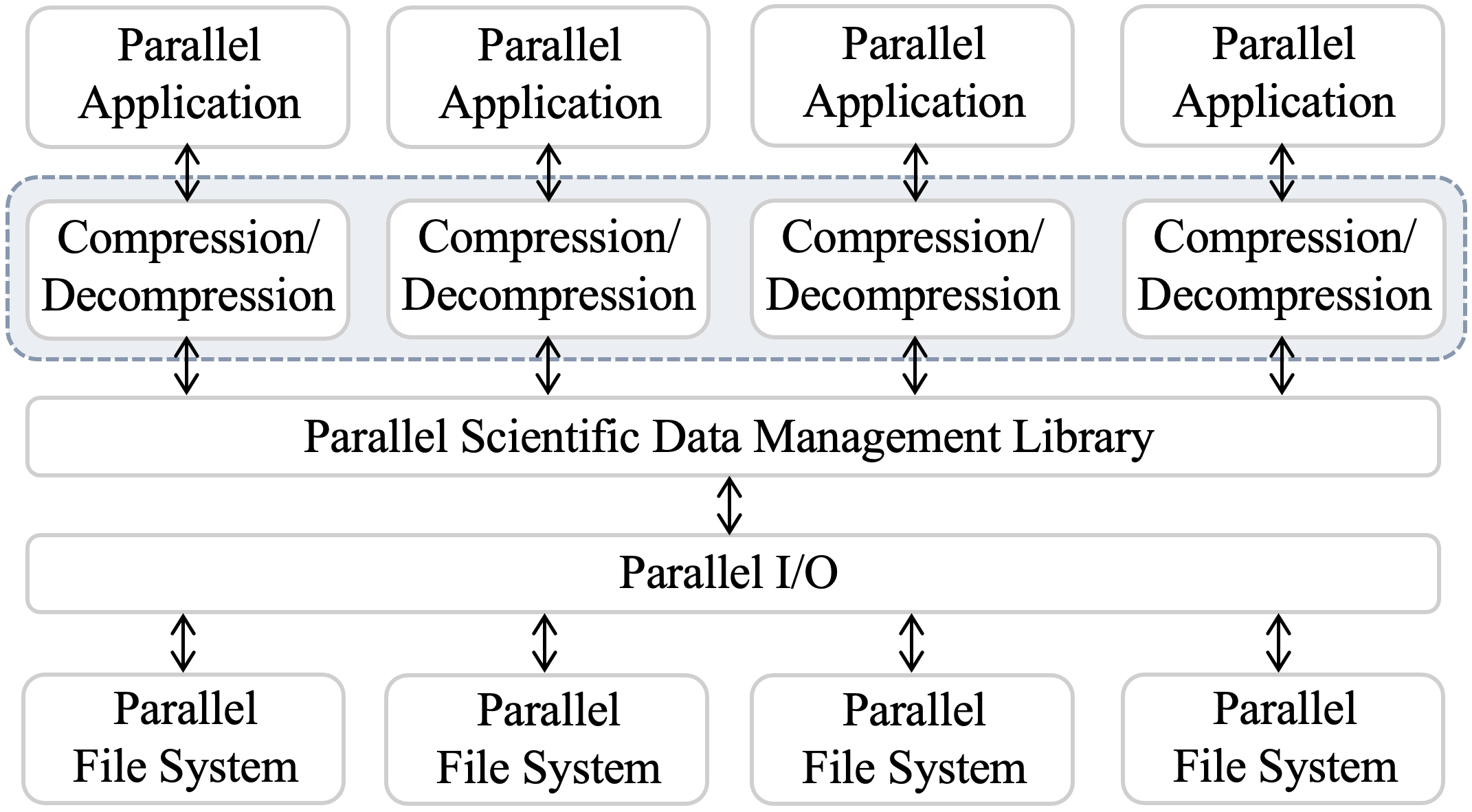}
%    \vspace{-7mm}
    \caption{Scientific data management with compression.}
    % \vspace{-2mm}
    \label{fig:fig-data-manage}
\end{figure}

\textcolor{black}{
In recent years, data management for scientific applications has become a fairly non-trivial challenge. Researchers must develop efficient data management approaches and software to handle extreme data sizes and unusual data movement characteristics. 
%and optimize specialized data management libraries due to the extreme size of the data and the unusual data movement . 
For example, the Advanced Photon Source (APS)~\cite{TheAdvan90:online} requires large datasets movement between the synchrotron and compute/storage facilities, and specialized data management libraries are proposed to handle such data movement~\cite{DataExch77:online}.
% the ATLAS Experiment~\cite{aad2008atlas} often requires moving large datasets between the storage facility and the computing facility, thus specialized data management libraries are proposed to handle such data movement~\cite{barisits2019rucio}.
%Special data models and their parallel data management libraries are also deeply coupled with scientific applications. 
In general, HDF5~\cite{hdf5}, netCDF \cite{li2003parallel}, and Adaptable IO System (ADIOS)~\cite{godoy2020adios} are 
%two of 
the most widely used data management software libraries for scientific applications running on high-performance computers. % providing efficient scientific data management in parallel.   
%that provides a parallel library for scientific data storage and post-hoc analysis.
However, these scientific data management techniques still suffer from extremely large datasets and subsequent I/O bottlenecks, therefore, compression techniques are often adopted by them.
Figure~\ref{fig:fig-data-manage} shows the abstraction of different layers in these data management systems with compression. %to overcome these issues. 
Note that compression functions as an individual layer in the data management system. Specifically, compression is performed between generating and storing the data, while decompression is performed between querying and deploying the data.
% is added to the system after generating/querying the data but before storing/deploying it, functioning as a layer to the data management software.
As a result, compression with high ratio can significantly improve the overall performance of large-scale scientific data management. % for large-scale scientific applications.
}

\textcolor{black}{
In this paper, considering HDF5 and its plugins \cite{HDF5Form1:online,pandasre24:online} are well received by the scientific community as a system supporting data management, we mainly focus our performance evaluation on HDF5 without loss of generality.
%\footnote{\textcolor{black}{Note that our method is independent of a specific data format and can be adapted to other data management systems.}}. 
Specifically, previous works propose several tools built on HDF5 that support querying. For example, Apache Drill~\cite{HDF5Form1:online} and Pandas~\cite{pandasre24:online} allow querying HDF5 metadata and data. 
% ICDE_revision % Moreover, prior research \cite{blanas2014parallel} also studies directly querying on HDF5 data.
Furthermore, its data reorganization has also been studied using FastBit indexes and transparently redirecting data accesses to the reorganized data or indexes~\cite{dong2013expediting}.
In addition, H5Z-SZ~\cite{hdf5filter-sz} provides a filter for integrating SZ into HDF5. 
Therefore, a deep understanding on lossy compressors can potentially significantly improve the overall performance of data management with HDF5. 
}

% Genral scientific data managment technique....

% software can use compression as filter ....

\subsection{Error-Bounded Lossy Compression}

% Floating-point data compression has been studied for decades and categorized into two types of compression: lossless compression and lossy compression. 
% Lossless compressors such as FPZIP~\cite{lindstrom2006fast} and FPC~\cite{FPC} keep the data intact but can only provide a compression ratio of about 2$\times$ on scientific data~\cite{son2014data}.
% Lossy compression, on the other hand, can compress data beyond lossless compression (typically one or more orders of magnitude) but introduces information loss in the reconstructed data.
%Compared to lossless compression, lossy compression can provide a much higher compression ratio w. 
\textcolor{black}{Lossy compression can compress data with extremely high compression ratio by losing non-critical information in the reconstructed data.
Two types of most important metrics to evaluate the performance of lossy compression are: (1) compression ratio, i.e., the ratio between original data size and compressed data size, or bit-rate, i.e., the number of bits on average for each data point on average (e.g., 32/64 for single/double-precision floating-point data before compression); and (2) data distortion metrics such as peak signal-to-noise ratio  (PSNR) to measure the reconstructed data quality compared to the original data.
}
In recent years, a new generation of high accuracy lossy compressors for scientific data have been proposed and developed for scientific floating-point data, such as SZ~\cite{di2016fast, tao2017significantly, liangerror} and ZFP~\cite{zfp}. These lossy compressors provide parameters that allow users to finely control the loss of information due to lossy compression. % maintaining sufficient information for scientific discovery
% \sout{Unlike traditional lossy compressors such as JPEG %~\cite{wallace1992jpeg} 
% which are designed for images (in integers), SZ and ZFP are designed to compress floating-point data and can provide a strict error-controlling scheme based on user's requirements.}
Generally, lossy compressors provide multiple compression modes, such as error-bounding mode. 
Error-bounding mode requires users to set an error type, such as the point-wise absolute error bound and point-wise relative error bound, and an error bound level (i.e., $10^{-3}$). The compressor ensures that the differences between the original data and the reconstructed data do not exceed the user-set error bound level.
% ICDE_revision % Fixed-rate mode means that users can set a target bit-rate (the number of bits to represent each data point), and the compressor guarantees the actual bit-rate of the compressed data to be lower than the user-set value. 

\textcolor{black}{In this paper, we mainly focus on ratio-quality modeling for prediction-based lossy compression.
The workflow of prediction-based lossy compression \cite{sz16,sz17,sz18,lindstrom2006fast} consists of three main stages: prediction, quantization, and encoding. 
First, each data point's value is predicted using a generic or specific prediction method. For example, the Lorenzo predictor can generally provide an accurate prediction for many simulation datasets  \cite{liang2018efficient,zhao2021optimizing,liu2021high}, while the spline interpolation based predictor can make a better prediction on seismic data (as proved in a recent study \cite{zhao2021optimizing}).
Then, each prediction error (the error between the predicted value and the original value) is quantized to an integer (called ``quantization code'') based on a user-set error mode and error bound. For instance, if the user needs to control the global upper bound of pointwise compression errors (the error between the original value and the reconstructed value), linear-scaling quantization scheme will be used with the quantization interval size \cite{sz17} equal to 2 times of the user-set error bound.
Lastly, one or multiple encoding techniques such as Huffman coding \cite{huffman1952method} (variable-length encoder) and LZ77 \cite{ziv1977universal} (dictionary encoder) will be applied to the quantization codes to reduce the data size.}

\textcolor{black}{Most previous works regarding lossy compression mainly aimed to improve the lossy compression ratio based on a specific algorithm. Specifically, some focus on how to improve the prediction efficiency and/or encoding efficiency~\cite{sz16,sz18,liang2018efficient,sz17,zhao2021optimizing}; some focus on application-specific optimizations based on empirical studies or trial-and-error methods~\cite{jin2020understanding,grosset2020foresight}. 
%previous works used empirical studies or trial-and-error methods for improving lossy compression in various scientific applications~\cite{cappello2019use,tao2019optimizing,fraz,sz16} and application-specific optimizations of lossy compression~\cite{jin2020understanding,grosset2020foresight}. 
In contrast, this paper is the first efficient attempt to provide an accurate estimation of compression ratio/quality and theoretically guide the use of lossy compression in database/scientific applications.}

Without an analytical model, existing lossy compression users have to use trial-and-error approach to obtain expected compression ratio and quality empirically.
\textcolor{black}{
% ICDE-revision % That is, one needs to experimentally run compression/decompression on the given dataset with a series of different compression configurations/parameters to measure the ratio-quality.
In other words, one needs to experimentally run compression and decompression on the given dataset with a series of different compression configurations (e.g., error bound) to measure the compression ratio/quality and generate the rate-distortion. Due to the high time cost, it is impossible to use this approach for in-situ optimization in database/scientific applications.
For example, a comprehensive framework, Foresight~\cite{grosset2020foresight}, has been developed to automate this process, but it is still limited to offline scenarios due to its high overhead~\cite{jin2020understanding}. Moreover, the offline optimization performs poorly in terms of compression-ratio/quality control over different data partitions or timesteps~\cite{jin2020adaptive}. 
}
%\sout{SZ -- a popular prediction-based lossy compressor, for example, cannot perform compression based on a given target compression ratio. Although ZFP \cite{zfp} provides a fix-rate mode (i.e., fix-compression-ratio), its compression quality is significantly worse than that of the error-bounding mode, as verified by Underwood et al. \cite{fraz}. This requires ZFP users to conduct a trial-and-error approach to obtain a target compression ratio given a scientific dataset~\cite{fraz,jin2020understanding}.
%%% Specifically, SZ is a prediction-based error-bounded lossy compressor for scientific data. SZ has three main steps: (1) predict each data point's value based on its neighboring points by using an adaptive, best-fit prediction method; (2) quantize the difference between the real value and predicted value based on the user-set error bound; and (3) apply a customized Huffman coding and lossless compression to achieve a higher ratio.
%In addition, the lack of compression ratio estimation and data quality estimation make lossy compression users rely on the trial-and-error technique for compression parameter optimization~\cite{jin2020understanding,grosset2020foresight}.}
In this paper, we build a systematic model for prediction-based lossy compression, supporting accurate and efficient ratio-quality estimation.

\subsection{Research Goals and Challenges}

\textcolor{black}{
Our work is the first work that provides a generic modeling approach for lossy compressors to accurately estimate its compression quality and hence avoid the trial-and-error overhead, which is an essential research issue for today's lossy compression work. 
}
% ICDE_revision % To provide a generalized, accurate lossy compression quality model with very low overhead, 
\textcolor{black}{ To achieve this,
four main challenges need to be addressed: 
(1) How to decompose prediction-based lossy compression into multiple stages and model the compression ratio for each stage?
We target to accomplish theoretical analysis for every compression stage (i.e., prediction, quantization and encoding) independently and propose the overall ratio-quality model based on them.
(2) How to reduce the time cost of extracting data information needed by the model? 
We target to design an efficient sampling strategy that can guarantee our prediction accuracy.
(3) How to model the quality degradation in terms of diverse post-analysis metrics?
We target to provide a guideline to incorporate new application-specific analysis metrics into our model by performing theoretical or empirical analysis.
(4) How does our model benefit real-world applications?
We target to design various optimization strategies for multiple use-cases to balance the compression ratio and the post-hoc analysis quality on reconstructed data.
}
\section{Ratio-Quality Modeling}
\label{sec:analysis}

In this section, we describe the overall design of our proposed ratio-quality model for prediction-based lossy compressors and present the detailed analysis of each component.

\subsection{Overall Design}
\label{sec:compressor}

\begin{figure}[t]
    \centering
    \includegraphics[width=1.0\linewidth]{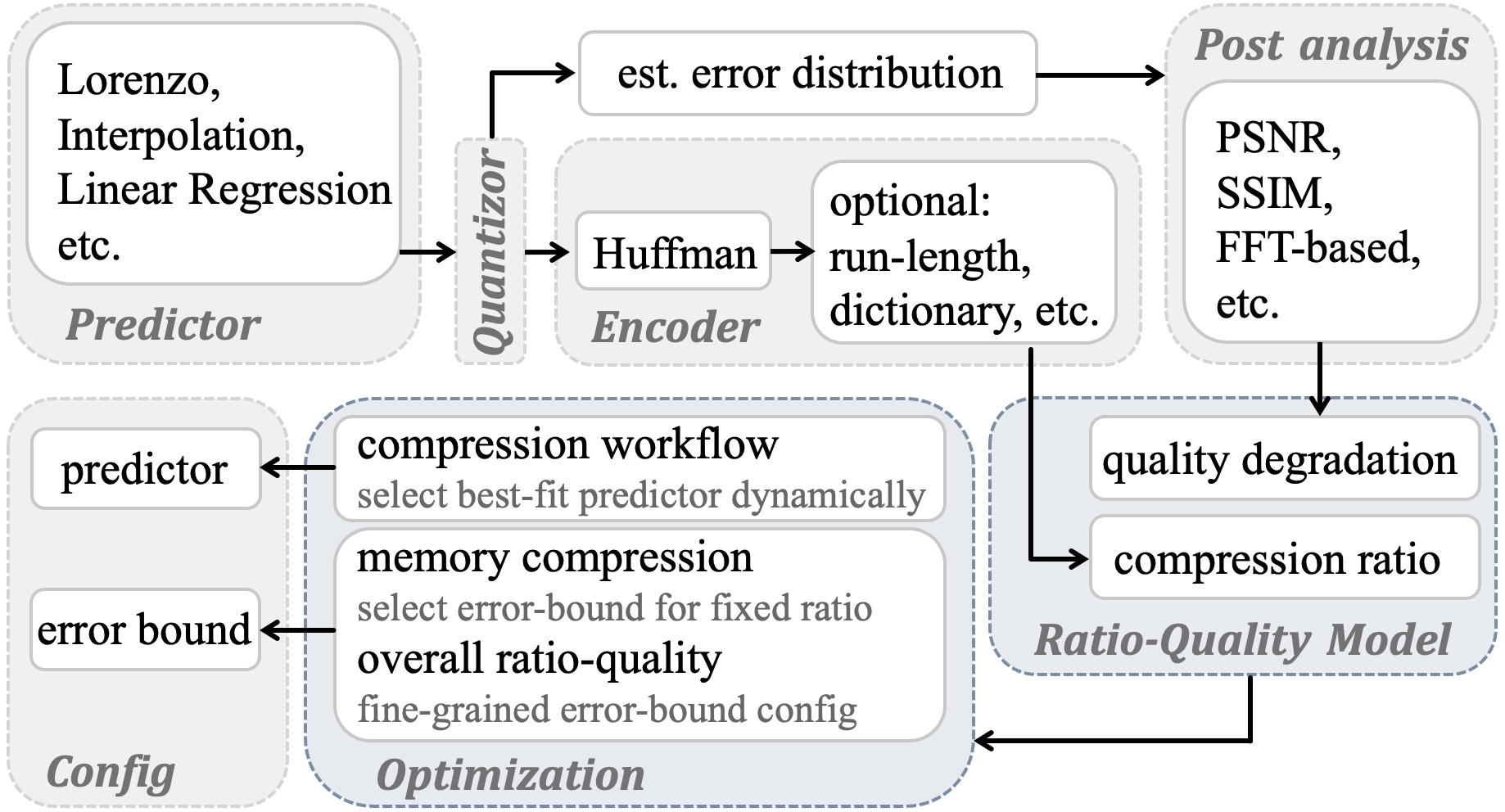}
%    \vspace{-7mm}
    \caption{An overview of ratio-quality modeling workflow for prediction-based lossy compression and scientific data analysis.}
    % \vspace{-2mm}
    \label{fig:fig-31-1}
\end{figure}

Figure~\ref{fig:fig-31-1} illustrates the workflow of our ratio-quality model, which is fully modularized with a high extensibility. 
Our ratio-quality model is built based on two main estimates: compression ratio and post-hoc analysis quality.
%%% (1) compression ratio estimation, which means we can provide an estimated bit-rate based on the compression configurations and the user-defined error bound; and (2) post-hoc analysis quality estimation based on the same input of ratio estimation. 
We build a ratio-quality model based on error bound and provide optimization for different predictors and error bounds. % for the prediction-based lossy compressors. 
% The detailed analysis for each module shown in Figure~\ref{fig:fig-31-1} can be found in Section~\ref{sec:design}.

To model compression ratio, we first examine the existing error-bound modes offered by lossy compressors. 
% we first determine the error-bound mode that is being used. This can provide us the information for quantization granularity and data pre-processing status. 
For example, a logarithm transformation is needed before compression for pointwise relative error bound mode \cite{liang2018efficient}. 
\textcolor{black}{Then, the main compression-ratio estimation consists of three modules for predictor, quantization, and encoder, respectively, as shown in Figure~\ref{fig:fig-31-1}.
We first model the predictor to provide an estimated histogram of prediction errors.
For example, a specifically designed sampling strategy is used for interpolation prediction.
After that, we model the quantization stage to estimate both error distribution and quantization-code histogram.
Finally, we estimate the compression ratio based on our theoretical analysis of the efficiencies of Huffman encoding and optional lossless compression.
}
%\sout{Then, we model the predictor to provide an estimated histogram of prediction errors. After that, we model the quantization section to provide estimates on both error distribution and quantization code histogram. Finally, we estimate the compression ratio based on our theoretical analysis for efficiency of Huffman encoding and optional lossless compression.}

To model the post-hoc analysis quality, we first determine the error distribution from the compressor quantization step. Then, we analyze the impact on post-hoc analysis quality by a theoretical derivation for error propagation based on hypothetical error injection to the datasets.
\textcolor{black}{For example, we predict the value of PSNR based on the variance of the estimated error distribution.}

In addition, our ratio-quality model can be leveraged for multiple use-cases, which will be described in detail in Section~\ref{sec:design}.
%%% (1) provides optimized compression predictor selection and error-bound mode selection based on the information extraction and the post-hoc analysis metrics of given dataset; 
% (2) provides accurate compression ratio estimation that can optimizes the compression performance under memory limited or memory controlled use cases; 
% and (3) provides in-situ compression optimization to balance between compression ratio and post-hoc analysis quality degradation, to achieve higher overall compression ratio with similar post-hoc analysis quality or post-hoc higher analysis quality with similar compression ratio compared to traditional trail-and-error approach.
Note that we provide a thorough modeling of multiple lossy compressor modules, while it is not necessary to apply the entire model for certain use-cases. The users, for example, can conduct memory compression based on a target ratio without post-hoc analysis quality estimation. This can minimize the computational overhead on demand. 
%In certain applications, we can further simplify some modules of the ratio-quality model to reduce the optimization overhead. %%%
We will detail our modularized model in the following sections. 
We follow three consecutive steps: (1) model compression ratio of popular encoders (i.e., Huffman encoder and run-length encoder); (2) refine compression ratio modeling for various predictors and quantizers; and (3) model quality degradation for both generic and specific post-hoc analysis.

\subsection{Modeling Encoder Efficiency}
\label{sec:encoder}

\begin{figure}[]
    \centering
    \includegraphics[width=0.97\linewidth]{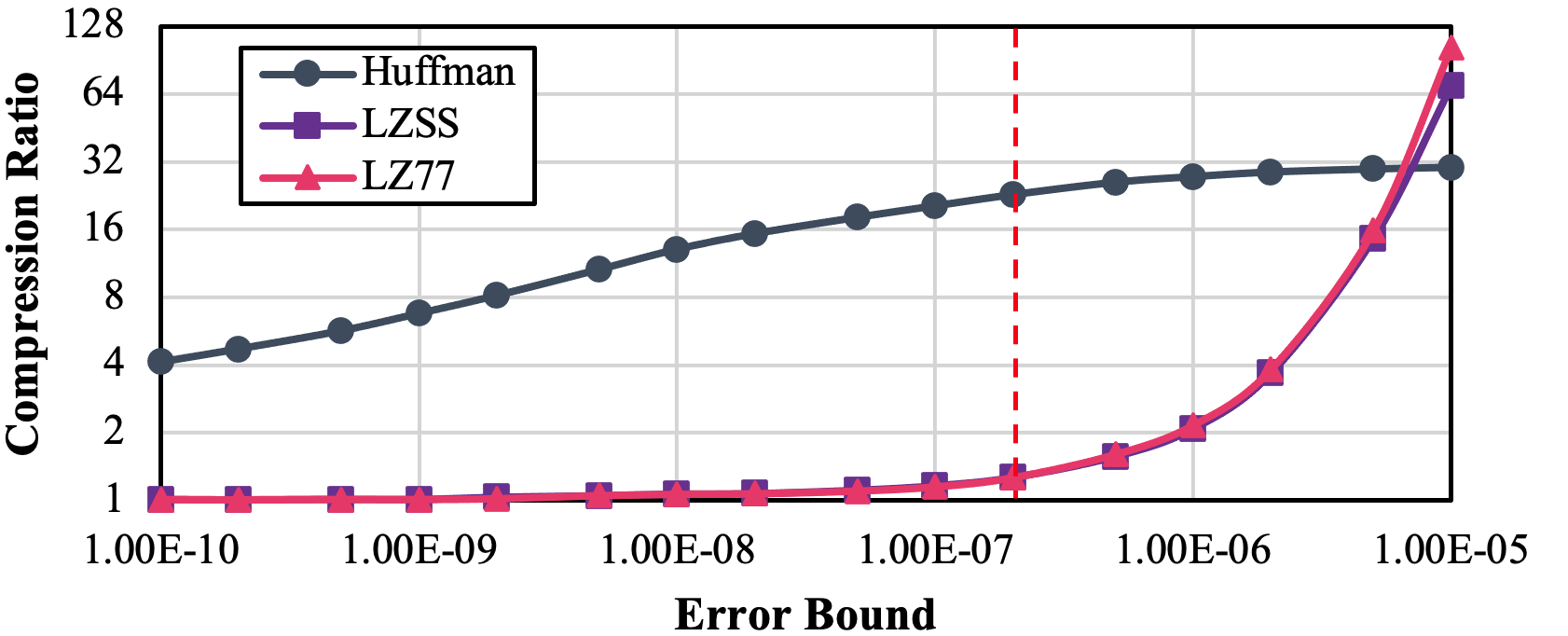}
%    \vspace{-3mm}
    \caption{Compression ratio from Huffman encoder and optional lossless encoder from Zstandard and Gzip on quantization code.}
    \label{fig:fig-4-1}
    % \vspace{-2mm}
\end{figure}

When encoding the quantization code, a Huffman encoder is applied first, followed by other optional lossless encoding techniques (e.g., run-length encoding, dictionary based encoding). Prior studies \cite{sz16,sz17,sz18,lindstrom2006fast} have pointed out that using the Huffman encoder for the quantization code plays a major role on the best overall encoding efficiency.
On the one hand, based on our experiments, we find the encoding efficiency provided by Huffman encoding is highly separated from the encoding efficiency provided by the optional lossless encoders, shown in Figure~\ref{fig:fig-4-1}. 
This means that the encoding efficiency provided by the optional lossless encoders only complements Huffman encoding after it reaches a certain limit ($\sim$1 bit per symbol).
On the other hand, we find that applying run-length encoding on the output of Huffman encoding can get the compression ratio very close to the one that uses an entire lossless compressor (e.g., Zstandard) after Huffman encoding. This is because the predictor always tries best to predict data points as accurately as possible, such that a large majority of the predicted values would fall within the error bound around the corresponding real values. The corresponding quantization code is marked as `zero' in this situation. Accordingly, zero would always dominate the Huffman codes, especially when the compression is performed with a relatively high error bound.
\textcolor{black}{We will demonstrate the effectiveness of this approximation approach in Section~\ref{sec:evaluation} (i.e., the ``Lossless Error'' column in Table~\ref{tab:Result}, where the lossless-encoding efficiency is predicted based on RLE only).}
% In fact, the quantization codes are almost independent random and the probability of efficiently representing consecutive non-zero codes is extremely low.
Thus, we estimate the encoder efficiency based on (i) Huffman encoding that encodes the frequency information and (ii) run-length encoding that encodes the spatial information (for high error bounds). Modeling each of these two coding methods includes two key compression modes: fix-error-bound (a.k.a., fix-accuracy) mode and fix-rate mode \cite{zfp-doc}, which will be detailed in the following discussion, respectively.

\subsubsection{Modeling Huffman Coding}

In what follows, we describe how to estimate the bit-rate (i.e., compression ratio) based on a given error bound first, and then derive the required error bound based on a target bit-rate.  

\noindent \textbf{Estimate bit-rate}:
We model the bit-rate $B$ which resulted from Huffman coding (i.e., the average bit length of each data point), in terms of the quantization codes that were generated from the previous steps (prediction + quantizer) as follows:
\begin{align}
 \small \textstyle
    B = \sum_{i=0}^{n}P(s_i)L(s_i) \approx -\sum_{i=0}^{n}P(s_i)\log_2P(s_i),
    \label{equ-1} 
\end{align}
where $n$ is the number of different Huffman codes, $P$ is the probability (or frequency) of given code $s_i$, $L$ is the length of given code $s_i$. We further represent the Huffman code length based on its probability with the binary base-2 numeral system. 
In Equation (\ref{equ-1}), when processing the code with the highest frequency, we need to adjust its bit-rate $-log_2P(s_i)$ to be the minimum code length (i.e., $1$ bit).
% Note that in our case, 1024 symbols are used for Huffman coding and thus are sufficient for this simplification. 

\noindent \textbf{Optimize error bound based on bit-rate:}
The optimized error bound $e^*$ for the target bit-rate $B^*$ based on the existing bit-rate $B$ is shown in the following equation:
\begin{align}
 \small\textstyle
    e^* = 2^{B - B^*}e,
    \label{equ-3} 
\end{align}
where $e$ is the profiled error bound with a bit-rate of $B$, by Equation (\ref{equ-1}).
We derive the above equation as follows. 
% \begin{theorem}
% The optimized error bound $eb^*$ for the target bit-rate $B'$ based on the existing bit-rate $B$ is shown in the following equation:
% \begin{align}
%  \small\textstyle
%     eb^* = 2^{B - B'}eb
%     \label{equ-3} 
% \end{align}
% \end{theorem}
\begin{prf}
Consider that a given error bound $e$ can provide a bit-rate of $B$, when doubling the error bound to $2e$, the quantization-code histogram also shrinks accordingly where the total number of symbols is reduced by $2\times$ and the probability (i.e., frequency) of each symbol would increase by $2\times$.
% \textcolor{red}{[Franck: possibility OR probability of occurence?]}
%%% Although the prediction-based compressor algorithm uses previous data points' quantized values to predict the value of current point that may affect the shape of quantization-code histogram. Based on our observation, this only applies for very large error bounds and hence few quantization bins. In our case, we simplify this to a fixed quantization-code histogram shape under different error bounds, thus a precise $2\times$ shrink when doubling the error bound.
In this case the bit-rate should be: %$B^\prime$ that equals to
\begin{align}
\label{equ-2} 
\begin{array}{l}
  \hspace{-3.7mm} B'\hspace{-0.5mm} = \hspace{-0.5mm}
    \hspace{-0.5mm}-\hspace{-0.5mm}\sum_{i=0}^{n/2}\hspace{-0.5mm}P'(s_i)\hspace{-0.5mm}\log_2\hspace{-0.5mm}P'(s_i)\hspace{-1mm}\approx\hspace{-1mm} -\hspace{-0.5mm}\sum_{i=0}^{n/2}\hspace{-0.5mm}2P(s_{2i})\hspace{-0.5mm}\log_22P(s_{2i}) \hspace{-4mm}\\
    \hspace{-1.7mm}=\hspace{-0.5mm} -\hspace{-0.5mm} \sum_{i=0}^{n/2}2P(s_{2i})\log_2\hspace{-0.5mm}P(s_{2i})\hspace{-0.8mm} -\hspace{-0.8mm} \sum_{i=0}^{n/2}2P(s_{2i})\log_22 \approx \hspace{-0.8mm}B\hspace{-0.8mm}-\hspace{-0.8mm}1.\hspace{-4mm}
\end{array}
\end{align}
%ICDE_revision%
% \begin{align}
% \label{equ-2} 
% % \begin{array}{l}
%   B' = 
%     -\sum_{i=0}^{n/2}P'(s_i)\log_2P'(s_i)\approx -\sum_{i=0}^{n/2}2P(s_{2i})\log_22P(s_{2i}) \notag \\
%     = - \sum_{i=0}^{n/2}2P(s_{2i})\log_2P(s_{2i}) - \sum_{i=0}^{n/2}2P(s_{2i})\log_22 \approx B-1.
% % \end{array}
% \end{align}
% \begin{align}
%  \footnotesize\textstyle
%   \hspace{-2mm} B'\hspace{-0.5mm} 
% = \hspace{-0.5mm}
%  \footnotesize\textstyle
%     \hspace{-0.5mm}-\hspace{-0.5mm}\sum_{i=0}^{n/2}\hspace{-0.5mm}P'(s_i)\hspace{-0.5mm}\log_2\hspace{-0.5mm}P'(s_i)\hspace{-1mm}\approx\hspace{-1mm} -\hspace{-0.5mm}\sum_{i=0}^{n/2}\hspace{-0.5mm}2P(s_{2i})\hspace{-0.5mm}\log_22P(s_{2i}) \notag\\
%              = 
%  \textstyle
%              \hspace{-0.5mm} -\hspace{-0.5mm} \sum_{i=0}^{n/2}2P(s_{2i})\log_2P(s_{2i}) + \sum_{i=0}^{n/2}2P(s_{2i})\log_22 \approx \hspace{-0.5mm}B\hspace{-0.5mm}-\hspace{-0.5mm}1\hspace{-0.5mm}
%             %  &\approx \sum_{i=0}^{n/2}(P(s_{2i-1})log_2P'(s_{2i-1})+P(s_{2i})log_2P'(s_{2i}))-1 \notag\\
%             %  &= B-1
%     \label{equ-2} 
% \end{align}
% Thus, we conclude that by doubling the error bound, the bit-rate with Huffman encoding decreases by $1$. %%%
Applying the above equation iteratively, we obtain the situation with any specific target bit-rate $B^*$ in Equation (\ref{equ-3}).
\end{prf}

% This is also consistent for continues error bound: if compression bit-rate is $B$ under error bound $eb$, then under new error bound $Neb$ the predicted bit-rate is $B'=B-log_2N$.
% Note that the SZ algorithm uses previous data points' quantized values to predict the value of current point based on Lorenzo prediction, which means different error bounds would affect the shape of quantization-code histogram. However, based on our experiments, this only applies for very large error bounds and hence few quantization bins. In our case, we simplify this to a fixed quantization-code histogram shape under different error bounds, thus a precise $2\times$ shrink when doubling the error bound.
% Similarly, we can compute the optimized error bound $eb^*$ for the target bit-rate $B'$ based on the existing bit-rate $B$ by the following equation:
% \begin{align}
%  \small\textstyle
%     eb^* = 2^{B - B'}eb
%     \label{equ-3} 
% \end{align}
% Note we can always provide an accurate Huffman efficiency estimation based on equition~\ref{equ-1}.
% The overall model for Huffman efficiency is the combination of both equation~\ref{equ-2} and the fitted model under high error-bound.

When the number of quantization bins is fairly small, the above estimation method is not applicable, because the approximation in Equation (\ref{equ-2}) no longer holds. 
% Based on our observations, w
We found Equation (\ref{equ-2}) starts to fall when the percentage of code zero (i.e., $p_0$) exceeds 50\% \textcolor{black}{based on our extensive experiments and datasets}. 
% \textcolor{red}{[Franck: I don't understand the next sentence]} 
In this case, we profile the histogram of quantization codes at $p_0=[0.5,0.8,0.95]$ and compute their corresponded $B$ from Equation~(\ref{equ-1}). Then, based on these pairs of $(p_0, B)$, we can interpolate a continuous function to provide a relationship between error bound and Huffman encoding efficiency.
Note that we profile the histogram at $p_0$ by keeping enlarging the width of the central bin of the histogram until its portion reaches $p_0$ where its width is $2e^*$.
% conduct three more computations at $p_0=[0.5,0.8,0.95]$ based on Equation (\ref{equ-1}) as anchor points to provide a consecutive function between error bound and Huffman efficiency.}

\subsubsection{Modeling Run-Length Encoding (RLE)}

As aforementioned, lossless encoders %for spacial information 
contribute to the compression ratio only when Huffman encoder reaches the limit, where zeros dominate the quantization codes. 
% For that situation, we model the optional lossless encoders with RLE. %run-length encoding (RLE).
% the encoded data array from Huffman encoder has two characteristic: (1) the data histogram is highly centralized and is zero dominated, assuming zero is used to encode the highest percentage of code in Huffman codebook; and (2) the data distribution is highly randomized due to data decorrelation by predictor in the previous process. Thus, we can model the run-length encoding efficiency on zeros to represent the lossless encoder efficiency for spacial information. Our experiment shows this modeling is accurate compared to using sufficient lossless encoder such as Gzip and Zstandard.
% \textcolor{black}{This is because the quantization codes (prediction errors) are decorrelated after effective prediction and contain almost no redundant information other than frequency information that can be efficiently reduced by Huffman coding; only when Huffman encoding reaches its limit under high error bounds, most of the quantization codes are coded as 1-bit zeros, providing additional continuity for further compression.}
% On the other hand, the design of the predictor defines a centralized quantization code distributed and zero always dominates the Huffman coding when reaching the limit.
Moreover, the quantization codes are independently random after an effective prediction due to its high decorrelation efficiency, causing an extremely low probability of consecutive non-zero codes.
%predictors' decorrelation and the probability of efficiently representing consecutive non-zero codes is extremely low.
Thus, we hereby model the RLE on zeros only.
% Thus, we model the efficiency of RLE by the average length of consecutive zeros and the length of representation for consecutive zeros:}
% \textcolor{red}{[Franck:why are we only considering consecutive zeros? Are we sure that no other symbols are consecutive?]} 

\noindent \textbf{Estimate compression ratio}: We model the compression ratio of RLE $R_{rle}$ by the following equation:
\begin{align}
 \small\textstyle
    R_{rle} = 1/(C_1(1-p_0)P_0 + (1-P_0)).
    \label{equ-run-4} 
\end{align}
Here $P_0$ is the percentage of footprint the code zero takes with respect to the full Huffman encoded data size, where $p_0$ is percentage of the number of zeros.
\begin{prf}
We first model the efficiency $E_0$ \textcolor{black}{(defined as the reciprocal of the reduction ratio)} for encoding zeros by the average length of consecutive zeros $n_0$ and the length $l_0$ of representation for consecutive zeros:
\begin{align}
 \small\textstyle
    E_{0} = C_1/n_0l_0,
    \label{equ-run-1} 
\end{align}
where $C_1$ is the fixed size of data to represent consecutive code. $l_0$ is 1 as the length for zero in Huffman codebook.
The overall compression ratio by RLE is:
\begin{align}
 \small\textstyle
    R_{rle} = 1/(E_{0}P_0 + (1-P_0)),
    \label{equ-run-2} 
\end{align}
considering that the data distribution is independent random as aforementioned, we have $n_0$ equal to:
\begin{align}
% \textstyle
    % n_0 &= 
\textstyle
    \sum_{n=1}^{\infty}np_0^{n-1}(1-p_0) = (1-p_0)\frac{d}{dp_0}\sum_{n=1}^{\infty}p_0^n= \frac{1}{1-p_0},
    % n_0 &= \sum_{n=1}^{\infty}np_0^{n-1}(1-p_0) = (1-p_0)\frac{d}{dp_0}\sum_{n=1}^{\infty}p_0^n = (1-p_0)\frac{d}{dp_0}\frac{1}{1-p_0} \notag\\
    % &= \frac{1}{1-p_0}
    \label{equ-run-3} 
\end{align}
From Equations (\ref{equ-run-1}), (\ref{equ-run-2}) and (\ref{equ-run-3}), we can get Equation (\ref{equ-run-4}).
\end{prf}

\noindent \textbf{Optimize error bound based on bit-rate:}
The optimized error bound $e^*$ is profiled from quantization code by the target percentage of zeros $p_0$ that is deduced from Equation~(\ref{equ-run-4}):
\begin{align}
\textstyle
    p_0 = \sqrt{1-R_{rle}^{-1}-((C_1-1)/2)^2} + (C_1-1)/2
    \label{equ-run-5} 
\end{align}
% We profile the width of the central bin in the quantization code histogram until its proportion reaches $p_0$, where its width is $2e^*$.
% The error bound $e^*$ is the width for quantization
Note we let $P_0 \approx p_0$ to derive Equation~(\ref{equ-run-5}) in the case of using RLE because code zero dominates the Huffman encoded data.

\subsection{Modeling Quantized Prediction Error Histogram}
\label{sec:predictor}

Quantuized prediction error histogram needs to be modeled, since the prediction-based lossy compression relies on an efficient predictor and quantizer to concentrate the input data information for high encoding efficiency. To this end, we calculate the distribution of prediction errors, based on which the quantization code histogram would be constructed for the encoder module. 
Note that the prediction-error distribution is different from the quantization-code distribution for the encoder module,
\textcolor{black}{where a highly accurate estimated distribution of prediction errors is demanded to estimate the distribution of quantization codes.}
In order to obtain accurate prediction error distribution with low overhead, we must apply suitable sampling strategy based on different predictors' design principles, to be detailed later.
For sparse scientific data, this step also determines the sparsity and removes the corresponding zeros in the prediction error distribution for getting high model accuracy.
We analyze and design the sampling strategy for all the three predictors used in SZ: Lorenzo, linear interpolation, and linear regression predictors. We set the sample rate always to 1\% to balance the accuracy and overhead based on our evaluation in Section~\ref{sec:evaluation}. Since our design is generic and modularized, new predictors and sampling methods can be added in the future work.

\subsubsection{Lorenzo Predictor}

The Lorenzo predictor~\cite{ibarria2003out} in SZ uses the previous few layers of data to build 1-2 levels of Lorenzo prediction for current value. 
% For example, a level-1 Lorenzo prediction on one dimensional data is to use previous two value to predict current value with linear regression.
In this case we randomly sample the given data and for each sampled point, then apply the Lorenzo predictor and collect the difference between predicted value and actual value. 
% Based on our experiment, we find a sampling rate of 3\% for Lorenzo predictor can provide us with high fidelity error distribution required by our quantization module.

\subsubsection{Linear Interpolation Predictor}

This predictor \cite{zhao2021optimizing} uses the surrounding data points to predict the current value. The prediction starts with the vertices of the input data to predict the middle point; then for each partition divided by this middle point, it performs the same procedure until reaching the smallest granularity \cite{zhao2021optimizing}. 
% In current implementation of SZ, it runs 1-D linear interpolation prediction on each data dimension by alternation for data dimensions higher than one. 
% One way to sample the data based on linear interpolation is to sample a block of data and perform the prediction on it. However, this can cause the sampled data contains only local information of the dataset that cannot represent the overall error distribution in some cases.
We propose to sample the data points randomly to extract the information from the entire dataset by considering different sampling rates in different interpolation levels. %\textcolor{black}{(i.e., the layer of interpolation performed).
% Since half of the values are predicted from the two neighboring data point, we use half of the sampled data to perform the distance $d=2^0$ linear interpolation prediction. 
% And for each larger granularity, we use half the rest sampled data to perform. 
Specifically, the sampling data in the current level is $2^{-n}$ than the previous level, where $n$ is the data dimension. 
% For each interpolation level one step larger than the last one, we use $2^{-n}$ of its sampling data, where $n$ is the data dimension. 
% \textcolor{red}{[Franck]:I don't understand your concept of granulatity.}
%%% For example, we use $1/8$ sampled data to perform the distance $d=2^2$ linear interpolation prediction. %%%
% % Based on our experiment, we find a sampling rate of 3\% for linear interpolation can provide us with high fidelity error distribution required by our quantization module.

\subsubsection{Linear Regression Predictor}

The linear regression predictor \cite{sz18} is performed by separating the data into small blocks (e.g., 6$\times$6$\times$ for 3D dataset) and uses a linear regression function to fit the data in each block. As it differs from the previous two predictors, we must sample the dataset by blocks to perform linear regression. 
% In current implementation of SZ, the block size is $8^n$, where $n$ is the data dimension. 
Thanks to the small block size used in SZ, performing sampling in the unit of block is able to represent the entire data for most scientific datasets even with a relatively low sampling rate.
% Based on our experiment, we find a sampling rate of 5\% for linear regression can provide us with high fidelity error distribution required by our quantization module.

\subsubsection{Quantization with Error Bound}
\label{sec:error-bound}

% Error-bounded lossy compressor (r.g., SZ, ZFP, ISABELA, FPZIP) provide various error-bound mode for user to actively define a strict error control scheme. In this paper, we focus on three error-bound modes that are widely used in today's scientific lossy compressors: Absolute Error Bound mode (ABS), Relative Error Bound mode (REL) and Power-Relative Error Bound mode (PW\_REL). 
% ABS mode let user to define an absolute error-bound and requests the reconstructed error to original data always remains within this absolute error-bound. 
% Similar to ABS mode, REL mode let user to define an relative error-bound and the compressor converts it to an absolute error-bound based on the value range of given dataset. 
% Different from both ABS and REL mode, PW\_REL mode let user to define an power relative error-bound and requests the reconstructed error to original data remain within the error-bound relative to each value.

In most of cases, we use the original value to perform the prediction in the sampling step instead of the reconstructed value used in the actual compression, since we observe that the error distribution differs little in between.
%This means the error distribution is slightly different from when performing the actual compression 
%Under most error bound situations, this difference is negligible based on our observations. 
Then, we quantize the sampled data based on certain error bounds to calculate an estimated quantization code histogram for the latter analysis such as modeling encoder efficiency.
% However, in order to further provide a correlation function between error bound and compression ratio for various ratio-quality modeling based optimization, we must combined this approach with our previous modeling for lossless encoders.
% More specifically, we conduct compression ratio estimation from few selected error bound as anchor points of Huffman encoder efficiency model. Our theoretical model for Huffman encoder efficiency is accurate until close the the limitation of 1 bit where more anchor points are contributed for building the correlation function.

For the situation with fairly high error bounds (which is determined by a threshold of $p_0$), the quantization code histogram estimated using original data values could suffer from large distortion.
\textcolor{black}{
For example, assume two points in a 1D array with the original values of [..., 0.0, 1.3] and the reconstructed values of [..., $0.5$, $0.5$] under the error bound of $1.0$ with the Lorenzo predictor, the prediction errors for the two points are [..., $0.0$, $1.3$] based on the original values, which fall into two different quantization bins with the bin size of $2.0$ (i.e., two times of the error bound); however, the prediction errors for the two points are [..., -$0.5$, $0.8$] based on the reconstructed values, which belong to the same bin.
%compared to that based on the reconstructed value is [..., -0.5, 0.8], which can result in the same bin in its histogram.
}
Thus we add a correction layer in our estimation. 
% Different Predictor result differently in terms of creating this uncertainty. 
% We mainly adjust this for the Lorenzo and the linear interpolation predictor. 
% For linear regression predictor, it uses a regression function to predict the data and the regression function parameters are compressed in only 1/10 of error bound compared to compressing the data value, result in little histogram changes even under very high error bound configuration. Thus, we apply no correction when predictor is set to linear regression.
% \textcolor{black}{We add a random bin transfer mechanism to our estimated quantization code histogram under very high error bound (i.e., when the highest percentage of single bin exceed 50\%).}
%%% Specifically, we let Lorenzo and linear interpolation predictors use the predicted value for next value prediction.
%instead of using the original value as we used for sampling. 
%%% Each bin of the histogram can transfer some quantization codes to a different bin to potentially fix this issue.
Specifically, when Lorenzo or linear interpolation predictor is used, we let each bin of the histogram transfer some quantization codes to a different bin to correct the estimation, making it close to using the lossy reconstructed value to predict the next point.
% \textcolor{black}{Each quantization code from the quantization code histogram based on sampled data has the potential of transfer to a different histogram bin to correct this.}
For linear interpolation, the theoretical maximum possible bin transfer of each quantization code is $\pm1$.
% \textcolor{red}{[Franck]: what is bin transfer?}
For Lorenzo, the number is $\pm7$.
%%% 1D, $\pm3$ for 2D, $\pm7$ for 3D data (bin size is the error bound).
However, we simplify the possible bin transfer to $\pm1$ for both predictors because of the extremely low possibility of higher cross bins transfer compared to $\pm1$, based on our observations.
% Also, centralized error distribution under high error bound reduces the possibility of bin switch and can be simplified represented by the percentage of the highest code in current quantization code histogram $p_0$. 
Thus, we propose to adjust the estimated histogram by transferring a certain number of codes between neighboring quantization bins, to simulate this uncertainty.
%%%caused by the discrepancy between the predictions based on raw and reconstructed values. 
We use the percentage of highest code in current quantization code histogram $p_0$ to model this transfer number, since it is highly related to the centralization of quantization codes and also fast to compute. 
The estimated number of codes in each bin will be evenly transferred to its neighboring bins, when the percentage of the most frequent code $p_0$ exceeds a threshold of 80\%.
% We provide an estimated number of codes in each bin that will be evenly transferred to its neighboring bins,when the percentage of the most frequent code, i.e.,p0(representing the centrality of the histogram),exceeds a threshold. For example, we plan to model the number of transferred codes as
% The lower $(1-p_0)$, the lower standard deviation of quantization code that more centralized.
We conclude to add the following random bin transfer when providing the histogram to the encoder module for this:
\begin{align}
\textstyle
    % N_{switch} = P_{switch}N = a(1-p_0)N
    N_{tran} = P_{tran} \cdot N = C_2 \cdot (1-p_0) \cdot N, \;  \; \text{when}  \; p_0 \geq \theta_2,
    \label{equ-quant-1} 
\end{align}
where $N_{trans}$ is the number of codes transferred from one bin to its neighboring bins, $C_2$ is an empirical parameter for the predictor based on our experiment, $N$ is the number of values in given bin. Specifically, $C_2=0.2$ for Lorenzo predictor and $C_2=0.1$ for linear interpolation predictor.

\subsection{Modeling Post-hoc Analysis Quality}

Post-hoc analysis quality is highly related to the analysis metrics used for specific scientific applications. In this work, we introduce two widely used analysis metrics: PSNR and SSIM. 
We first provide an estimated error distribution of reconstructed data.
Then, we provide a theoretical analysis on each of the analysis metrics by propagating the compression error distribution function in the metric computation.
% \textcolor{red}{[Franck]: I don't understand the following sentence} 
For more domain-specific analysis metrics, the same principle can be adopted to build the post-hoc analysis quality model.
We also provide a guideline in the following sections.

\subsubsection{Error Distribution}

% \begin{figure}[]
%     \centering
%     \includegraphics[width=0.95\linewidth]{Figures/error-dis.png}
%     % \vspace{-2mm}
%     \caption{Error distribution of SZ lossy compression with different error bound. Experimented on Nyx dataset.}
%     \label{fig:fig-4-2}
% %    \vspace{-4mm}
% \end{figure}

We first provide the average and variance of the error distribution that are used for the error propagation analysis.
Error distribution of the reconstructed data is determined by the user defined error bound mode and the error bound value. In most of the cases, the error distribution of prediction-based lossy compressors forms a uniform distribution.
In which case $\mu(E) = 0$, and we have:
\begin{align}
% \textstyle
%     \mu(E) &= 0 \notag\\
    \sigma(E)^2 &= 
\textstyle
    \sum_{i=0}^N(E[i]^2-\mu^2) \approx \int_{-e}^{e}\frac{1}{2e}x^2dx = \frac{1}{3}e^2
    \label{equ-error-1} 
\end{align}
Here $e$ is the error bound. 
However, under high error bounds, we observe that the error distribution combines uniform distribution and centralized distribution.
% shown in Figure~\ref{fig:fig-4-2}.
This is because when the quantization bin size is fairly large, the error distribution will contain both a near-uniform distribution from non-central bins and a centralized distribution from the central bin.
% This is because the quantization bin size (i.e., the error-bound) is too large and cause the prediction error distribution to appear in the reconstructed error distribution, which can be considered as the combination of the distribution of each quantization bin combines.
% When error bound is very large, the error distribution of central bin in the quantization code histogram have higher weight in terms of reflecting the overall error distribution. 
More specifically, the weight is the percentage of the central bin in the quantization code histogram $p_0$, which is also the highest bin. Thus, we can separate the values within the central bin and others to have (with $\mu(E) = 0$):
\begin{align}
%  \small\textstyle
%     \mu(E) &= 0 \notag\\
\textstyle
    \sigma(E)^2 &= 
\textstyle
    \sum_{i=0}^{(1-p_0)N}(E[i]^2-\mu^2) + \sum_{i=0}^{p_0N}(E[i]^2-\mu^2) \notag\\
% \textstyle
    &= \textstyle (1-p_0)\frac{1}{3}e^2 + p_0\sigma(B[0]),
    \label{equ-error-2} 
\end{align}
where $\sigma(B[0])$ is the variance of values inside the central bin $B[0]$, which can be computed by our sampled data from the predictor module.
% This error distribution change must be considered for some post-hoc analysis (i.e., PSNR for Hurricane data). %%%

\subsubsection{Modeling PSNR}
We model the PSNR of reconstructed-original data as follows:
\begin{align}
% \begin{array}{l}
 \textstyle
    {PSNR}(D',D) = 20 \log_{10}(minmax) - 10\log_{10}(\sigma(E)^2)
    % \end{array}
    \label{equ-psnr-2} 
\end{align}
where $D'$ and $D$ is the reconstructed data and original data, respectively, and $minmax$ is the value range.
\begin{prf}
We start with modeling the mean squared error (MSE) based on static error distribution (i.e., uniform distribution).
The MSE between reconstructed data and original data equals to the variance of the compression error distribution:
\begin{align}
 \small\textstyle
    MSE(D', D) &= 
 \small\textstyle
    \sum_{i=0}^N(D'[i]^2 - D[i]^2)  \notag \\
    &= 
 \small\textstyle
    \sum_{i=0}^N(E[i]^2-\mu^2) = \sigma(E)
    \label{equ-psnr-1} 
\end{align}
where $E$ is the error distribution, $\mu$ is the average of $E$ that is usually zero from our evaluated predictors. Then, we compute the estimated PSNR by:
\begin{align}
 \small\textstyle
    PSNR(D',D) &= 10\log_{10} \left(\frac{minmax^2}{MSE} \right)
    % &= 20 \log_{10}(minmax) - 10\log_{10}(\sigma(E)^2)
    \label{equ-psnr-3} 
\end{align}
The above equation can deduce to Equation~(\ref{equ-psnr-2}).
\end{prf}

% We start the error propagation modeling for Peak Signal-to-Noise Ratio (PSNR) by modeling the mean squared error (MSE) with static error distribution (i.e., uniform distribution). Then, the PSNR estimation is performed from the MSE.
% The MSE between reconstructed data and original data equals to the variance of the compression error distribution:
% \begin{align}
%  \small\textstyle
%     MSE(D', D) &= 
%  \small\textstyle
%     \sum_{i=0}^N(D'[i]^2 - D[i]^2)  \\
%     &= 
%  \small\textstyle
%     \sum_{i=0}^N(E[i]^2-\mu^2) = \sigma(E)
%     \label{equ-psnr-1} 
% \end{align}
% Where $D'$ is the reconstructed data, $D$ is the original data, $E$ is the error distribution, $\mu$ is the average of $E$ that is usually zero from our evaluated predictors. Then, we compute the estimated PSNR by:
% \begin{align}
%  \small\textstyle
%     PSNR(D',D) &= 10\log_{10}(\textstyle\frac{minmax^2}{MSE}) \notag \\
%     &= 20 \log_{10}(minmax) - 10\log_{10}(\sigma(E)^2)
%     \label{equ-psnr-2} 
% \end{align}
% Where $minmax$ is the value range of given dataset, and $\sigma(E)$ is provided from previous error distribution analysis.

\subsubsection{Modeling SSIM}

We model the Structural SIMilarity index (SSIM) of reconstructed-original data as follows:
\begin{align}
 \small\textstyle
    % SSIM(D', D) &=\frac{2\sigma_{D}^2 + 2\sum_{i=0}^N(D[i]-\mu_D)E[i]+C_2}{2\sigma_{D}^2 + 2\sum_{i=0}^N(D[i]-\mu_D)E[i]+C_2+\sigma(E)^2} \notag\\
    SSIM(D', D) &=\frac{2\sigma_{D}^2+C_3}{2\sigma_{D}^2 +C_3+\sigma(E)^2} 
    % &= \frac{2\sigma_{D}^2 + }{}
    \label{equ-ssim-5} 
\end{align}
Where $C_3$ is a constant parameter when computing SSIM.
\begin{prf}
We also propagate the error distribution function in the computation of the SSIM.
\begin{align}
%  \small\textstyle
    % SSIM(D', D) &= \frac{2\mu_{D'}\mu_{D}+C_1}{\mu_{D'}^2+\mu_{D}^2+C_1} \frac{2\sigma_{D'}\sigma_{D}+C_2}{\sigma_{D'}^2+\sigma_{D}^2+C_2} \frac{\sigma_{D'D}+C_3}{\sigma_{D'}\sigma_{D}+C_3} \notag\\
    SSIM(D', D)=\frac{(2\mu_{D'}\mu_{D}+C_4)(2\sigma_{D'D}+C_3)}{(\mu_{D'}^2+\mu_{D}^2+C_4)(\sigma_{D'}^4+\sigma_{D}^2+C_3)}
    \label{equ-ssim-1} 
\end{align}
Here $C_3, C_4$ are both constant values. Considering the error distribution of reconstructed data from the original data, we assume $\mu(E) = 0$ on a large number of values in our error propagation analysis:
\begin{align}
%  \small\textstyle
    SSIM(D', D) &=\frac{2\sigma_{D'D}+C_3}{\sigma_{D'}^2+\sigma_{D}^2+C_3}
    \label{equ-ssim-2} 
\end{align}
For the variance of reconstructed data $\sigma_{D'}$, we have:
\begin{align}
 \small \textstyle
    \sigma_{D'}^2 &= \textstyle\sum_{i=0}^N((D[i]+E[i]) - \mu_D)^2 \approx \sigma_{D}^2 + \sigma(E)^2
    % &= \sigma_{D}^2 + 2\sum_{i=0}^N(D[i]-\mu_D)E[i] + \sigma(E)^2 \approx \sigma_{D}^2 + \sigma(E)^2
    \label{equ-ssim-3} 
\end{align}
Similarly, for the covariance $\sigma_{D'}^2$ between the reconstructed data and the original data, we have:
\begin{align}
 \small\textstyle
    \sigma_{D'D} &= \textstyle\sum_{i=0}^N((D[i] - \mu_D)((D[i]+E[i]) - \mu_D)) \notag\\
    &= \sigma_{D}^2 + \textstyle\sum_{i=0}^N(D[i]-\mu_D)E[i] \approx \sigma_{D}^2
    \label{equ-ssim-4} 
\end{align}
Based on Equations (\ref{equ-ssim-1}), (\ref{equ-ssim-2}) and (\ref{equ-ssim-3}), we can get Equation (\ref{equ-ssim-5}).
% \begin{align}
%  \small\textstyle
%     % SSIM(D', D) &=\frac{2\sigma_{D}^2 + 2\sum_{i=0}^N(D[i]-\mu_D)E[i]+C_2}{2\sigma_{D}^2 + 2\sum_{i=0}^N(D[i]-\mu_D)E[i]+C_2+\sigma(E)^2} \notag\\
%     SSIM(D', D) &=\textstyle\frac{2\sigma_{D}^2+C_2}{2\sigma_{D}^2 +C_2+\sigma(E)^2} 
%     % &= \frac{2\sigma_{D}^2 + }{}
%     \label{equ-ssim-5} 
% \end{align}
% Where all parameters can be provided by original data and our error distribution analysis.
Note that we simplify one of the terms in Equations (\ref{equ-ssim-3}) and (\ref{equ-ssim-4}) to its expected value 0 because this term would also show on both sides of the fraction in Equation (\ref{equ-ssim-5}) and its variance has little impact on the overall estimation.
% \textcolor{red}{[Franck]: what is "they"?}
% Sian: Refereed to the the term we simplified in equ-3 and 4, added some clarification.
\end{prf}

\subsubsection{Data Specific Post-hoc Analysis}

Specifically designed analysis metrics are also used for some scientific dataset, such as the Power Spectrum analysis and Halo Finder analysis used for Nyx dataset to identify the halo distribution. Previous research has performed error propagation analysis for FFT based Power Spectrum and Halo Finder with given error distribution~\cite{jin2020adaptive}. 
However, it uses uniform error distribution when modeling the analysis quality that can result differently under high error bound range. 
By using our newly proposed error distribution model, we can further improve the estimation accuracy of FFT-based analysis that is evaluated in Section~\ref{sec:evaluation}. 
A general guideline to quantify the degradation of domain-specific post-hoc analysis on reconstructed data is similar to our analysis process for PSNR and SSIM; we can adapt the post-hoc analysis computation to include the estimated compression error distribution function.

% \subsection{ratio-quality Model}

% With our proposed modeling for error-bound to bit-rate estimation function and modeling for error-bound to analysis quality degradation estimation function. we can combined the two into the ratio-quality Model, which not only can allow us to estimate the compression ratio and analysis quality with given compression configuration without computational consuming trail-and-error, but also can estimate the compression ratio by acceptable analysis quality degradation level and vice versa. Moreover, the ratio-quality Model can provide the slope between compression ratio and analysis quality, which can be utilized when balancing the compression strategy over multiple data partitions.

\section{Use-Cases of the Ratio-Quality Model}
\label{sec:design}

In this section, we introduce three use-cases leveraging our ratio-quality model to significantly improve the performance of prediction-based lossy compressors. 
% Including predictor optimization, memory compression optimization and fine-grained ratio-quality optimization.

\subsection{Compression Predictor Selection}

% \begin{figure}[]
%     \centering
%     \includegraphics[width=1.0\linewidth]{Figures/PSNR-Rate.png}
%     % \vspace{-2mm}
%     \caption{The PSNR to bit-rate curve of multiple predictors with different error-bound. Evaluated with ARAMCO dataset.}
%     \label{fig:fig-32-1}
% %    \vspace{-4mm}
% \end{figure}

The first use-case of our proposed ratio-quality model is to adaptively select the best predictor for any dataset and error bound.
In general, using larger error bounds usually results in higher compression ratios and lower data quality. 
However, the efficiency of each predictor differs in a certain range of ratio-quality curves.
% The efficiency of bilinear interpolation compared with Lorenzo prediction is higher when bit-rate is under 2 bits but lower above 2 bits. This indicate the most optimal solution would be using bilinear interpolation when bit-rate is under 2 bits and using Lorenzo prediction when above 2 bits.
With our proposed model, we can provide ratio-quality estimations for each predictor and select the best-fit predictor for a given error bound or target ratio. 
Compared to the existing methods \cite{tao2019optimizing,liang2019improving} that use the trial-and-error approach and sample prediction error for every error bound to select the optimal configuration, our method can help significantly reduce the optimization overhead with one-time sampling and efficient estimation.
Moreover, the existing methods make decisions only based on the compression ratio, whereas ours considers both ratio and post-analysis quality. 

% Based on our evaluation, the performance improvement of predictor optimization is highly effective for numbers of scientific datasets, especially for smooth-transition datasets (e.g., Hurricane, ARAMCO). On which the linear interpolation predictor shows significantly higher performance compared to Lorenzo predictor under low bit-rate scenario while shows worse performance under high bit-rate scenario. Our model can efficiently identify the critical point to switch between the two predictors, results in high compression performance for all bit-rate range with significantly lower overhead compared to traditional trail-and-error approach.

\subsection{Memory Compression with Target Ratio}

For applications that stores compressed data in memory and require a specific maximum footprint, our model estimates the compression ratio of any given dataset, which can provide an optimization strategy to efficiently utilize available memory.
% For example, when compressing the activation data for training deep neural networks with error-bounded lossy compressor, the compressed activation data is temporally stored on limited GPU memory and ready for reconstruction when needed~\cite{jin2021novel}. %%%
For these memory compression optimizations, we provide two optimization strategies.
First, if the application is not strictly limiting the bit-rate, such as saving compressed data on GPUs where the spilled data can be migrated to CPU, we create a target bit-rate for given one or multiple datasets that is 20\% lower than the limitation to allow uncertainty between estimation and real compression. We then compress the data with optimization towards the bit-rate target and allow data movements when exceeding the bit-rate limitation. This expensive but rare situation introduces little overhead to the system thanks to the high accuracy of our compression ratio modeling and the slightly lower target bit-rate for optimization.
Second, if the application strictly limits the bit-rate, we can also apply a similar strategy to provide a first-round optimization. Then, for rare situations where the actual size of compressed data still exceeds the limitation, we adjust the second-round optimization and re-compress the data to prevent overflows.

\subsection{In-Situ Compression Optimization}

When applying error-bounded lossy compression to a scientific dataset, one of the most common requested optimization is to balance between the compression ratio and reconstructed data quality. 
For dataset that is considered as a combination of multiple partitions, we are able to characterize each partition specifically based on a number of metrics (e.g., post-hoc analysis used and certain local data information extraction) which would then be used to decide which compression configuration to apply. Thus, we can optimize the compression performance individually for each partition with overall compression ratio and overall analysis quality as ojectives. 
Note the data partitions referred to as the partition of the entire data used for post-hoc analysis, such as data on multiple ranks or have multiple timesteps for post-hoc analysis.
% Even for dataset that is considered as a single unit, we can still apply the ratio-quality model to optimize the compression configuration with significantly lower overhead compared to traditional approach.
Such optimization is infeasible by previous solutions because of the exponentially increasing combinations for trial-and-error with increasing number of partitions.

% In this paper, we present a case study for RTM data. The data used for analysis is considered as a stacking image of data from multiple timesteps. The proposed optimization strategy is used to balance the ratio-quality for data from each partition. We demonstrate its performance in Section~\ref{sec:evaluation}.

\begin{table}[]
% \vspace{3mm}
\renewcommand*{\arraystretch}{1.4}
\centering
\ttfamily
\scriptsize
\caption{Details of Tested Datasets}
%\vspace{-4mm}
\newcommand\alignmiddle[2]{
\makebox[3em][r]{$(#1$}\makebox[.8em]{$,\ $}\makebox[2em][l]{$#2)$}}

% \begin{tabular}{@{} c c | l c @{}}
%     % \hline
% \toprule
%     Dimension
%      & Size
%      & Field
%      & Value Range
%     \\ 
%     % \hline
% \midrule
%     \multirow{4}{*}{
%         \begin{tabular}[c]{@{}c@{}}
%             $\phantom{0}512\times\phantom{0}512\times\phantom{0}512$    \\
%             $1024\times1024\times1024$ \\
%             $2048\times2048\times2048$
%         \end{tabular}
%     }
%      & \multirow{4}{*}{
%         \begin{tabular}[c]{@{}r@{}}
%             6.6 GB \\ 52 GB \\ 352 GB
%         \end{tabular}
%     }
%      & Baryon Density
%      & \alignmiddle{0}{10^5}%$(0, 10^5)$
%     \\ 
%     %\cline{3-4}
%      &
%      & Dark Matter Density
%      & \alignmiddle{0}{10^4} %$(0, 10^4)$
%     \\ 
%     %\cline{3-4}
%      &
%      & Temperature
%      & \alignmiddle{10^2}{10^7} %$(10^2, 10^7)$
%     \\ 
%     %\cline{3-4}
%      &
%      & Velocity
%      & \alignmiddle{-10^8}{10^8} %$(-10^8, 10^8)$
%     \\ 
%     % \hline
%     \bottomrule
% \end{tabular}

% \begin{table}[]

\newcommand{\SUPERBOLD}{\fontfamily{ugq}\selectfont}

\resizebox{\linewidth}{!}{
\begin{tabular}{@{} l|c|c|c|c @{}}
\hline
\SUPERBOLD Name      & \SUPERBOLD Dim     & \SUPERBOLD Size   & \SUPERBOLD Description & \SUPERBOLD Format          \\ \hline
CESM~\cite{cesm-atm}      & 2D     & 1.47GB & Climate simulation & NetCDF \cite{rew1990netcdf}     \\ \hline
% EXAALT    & 1D       & 60MB   & Molecular dynamics simulation    \\ \hline
EXAFEL~\cite{lcls}    & 4D & 51MB   & Instrument imaging   & HDF5 \cite{hdf5}       \\ \hline
Hurricane~\cite{hurricane} & 3D   & 1.25GB & Weather simulation    & Binary           \\ \hline
HACC~\cite{hacc}      & 1D     & 19GB   & Cosmology simulation     & GIO~\cite{gio}        \\ \hline
Nyx~\cite{nyx}       & 3D   & 2.7GB  & Cosmology simulation     &HDF5        \\ \hline
% NWChem    & 1D     & 16GB   & Two-electron repulsion integrals \\ \hline
SCALE~\cite{scale}     & 3D  & 4.9GB  & Climate simulation       & NetCDF       \\ \hline
QMCPACK~\cite{qmcpack}   & 3D     & 1GB    & Atoms' structure  & HDF5  \\ \hline
Miranda~\cite{miranda}   & 3D   & 1.87GB & Turbulence simulation     & Binary      \\ \hline
Brown~\cite{sdrbench}     & 1D       & 256MB  & Synthetic Brown data    & Binary                \\ \hline
RTM~\cite{rtm}       & 3D   & 682GB  & Reverse time migration     & HDF5      \\ \hline
\end{tabular}
}
% \end{table}

% Name      & Dimension     & Size   & Description                      \\ \hline
% CESM      & 1800x3600     & 1.47GB & Climate simulation               \\ \hline
% EXAALT    & 2869440       & 60MB   & Molecular dynamics simulation    \\ \hline
% EXAFEL    & 10x32x185x388 & 51MB   & LCLS instrument images           \\ \hline
% Hurricane & 100x500x500   & 1.25GB & Weather simulation               \\ \hline
% HACC      & 280953867     & 19GB   & Cosmology simulation             \\ \hline
% Nyx       & 512x512x512   & 2.7GB  & Cosmology simulation             \\ \hline
% NWChem    & 102953248     & 16GB   & Two-electron repulsion integrals \\ \hline
% SCALE     & 98x1200x1200  & 4.9GB  & Climate simulation               \\ \hline
% QMCPACK   & 69x69x115     & 1GB    & Electronic structure of atoms    \\ \hline
% Miranda   & 256x384x384   & 1.87GB & Turbulence simulations           \\ \hline
% Brown     & 8388609       & 256MB  & Synthetic                        \\ \hline
% RTM       & 235x449x449   & 682GB  & Reverse time migration           \\ \hline
% \vspace{-4.5mm}
\label{tab:DataDetail}
% \vspace{-2mm}
\end{table}

% \begin{table}[]
% % \vspace{3mm}
% \renewcommand*{\arraystretch}{1.3}
% \centering
% \ttfamily
% \scriptsize
% \caption{Details of Tested Datasets}
% %\vspace{-4mm}
% \input{Tables/tab-1}
% % \vspace{-4.5mm}
% \label{tab:DataDetail}
% % \vspace{-2mm}
% \end{table}

\section{Experimental Evaluation}
\label{sec:evaluation}

In this section, we present the evaluation results of our proposed ratio-quality model for prediction-based lossy compressors. We compare our approach with previous strategies in terms of both accuracy and performance. Next, we evaluate our model on different use-cases that can significantly improve the compression performance. 

\subsection{Evaluation Setup}

We perform our evaluation with the SZ3 \cite{sz3}, which is a modularized prediction-based lossy compression framework. 
%Considering our workflow can naturally scale up due to no inter-node communication, we conduct our experiment on a single node from the Frontera system \cite{frontera} at TACC, which is equipped with two 28-core Intel Xeon Platinum 8280 CPUs.
%and its  Longhorn subsystem \cite{longhorn}
\textcolor{black}{We conduct our experiments on the Bebop cluster \cite{BebopLab56:online} at Argonne, each node is equipped with two 18-core Intel Xeon E5-2695v4 CPUs and 128GB DDR4 memory. Considering that our workflow can naturally scale up due to no inter-node communication, we evaluate our model accuracy and use-cases study on a single node and the parallel data management performance on 8 nodes with 128 CPU cores.}
Moreover, we use 10 real-world scientific datasets from the Scientific Data Reduction Benchmarks \cite{sdrbench} in the evaluation. Table~\ref{tab:DataDetail} shows the detail of our tested datasets.

\subsection{Accuracy of Compression Ratio Model}

We first evaluate the accuracy of our modeled compression ratio. Based on our analysis in Section~\ref{sec:compressor}, we first evaluate the accuracy of our sampled prediction error from the predictor module. Then, we conduct our evaluation with two encoder setup situations: (1) we encode the quantization code with Huffman encoder only, and (2) we encode the quantization code with both Huffman encoder and an optional Lossless encoder, in which case we use Zstandard to measure the actual compression ratio in this paper. 
% \textcolor{red}{[Franck]: in the previous sections you focused and modeled run length encoding. Why are you considering Zstandard for the evaluation? Why your model of run length has any chance to represent Zstandard?}
% \textcolor{black}{[Dingwen]: Hi Franck, this point is a very good point worth explaining/emphasizing multiple times in the paper. We use RLE to model the general lossless encoders based on our assumption: zeros are dominant when ratio is very high (bitrate close to 1), other consecutive nonzeros (in an extremely low probably) have little contribution to the ratio. So, in the evaluation, we can evaluate Zstandard or any other lossless encoders and check our modeling accuracy based on RLE.  Probably, we should show a figure to compare RLE and Zstandard under high ratios to prove our assumption.}
Note that as aforementioned Section~\ref{sec:encoder}, we model the efficiency of optional lossless encoder based on RLE regardless of which lossless encoder is used, since the quantization codes for lossless encoding are highly decorrelated and zero-dominated.

\subsubsection{Sampled Prediction Error}

% \begin{table*}[]
% % \vspace{3mm}
% \renewcommand*{\arraystretch}{1.3}
% \centering
% \ttfamily\scriptsize
% \caption{Details of Evaluation Results on Tested Data and Fields}
% %\vspace{-4mm}
% \input{Tables/tab-2}
% \vspace{-4mm}
% \label{tab:Result}
% \end{table*}

\begin{figure}[]
    \centering
    \vspace{-2mm}
    \includegraphics[width=0.93\linewidth]{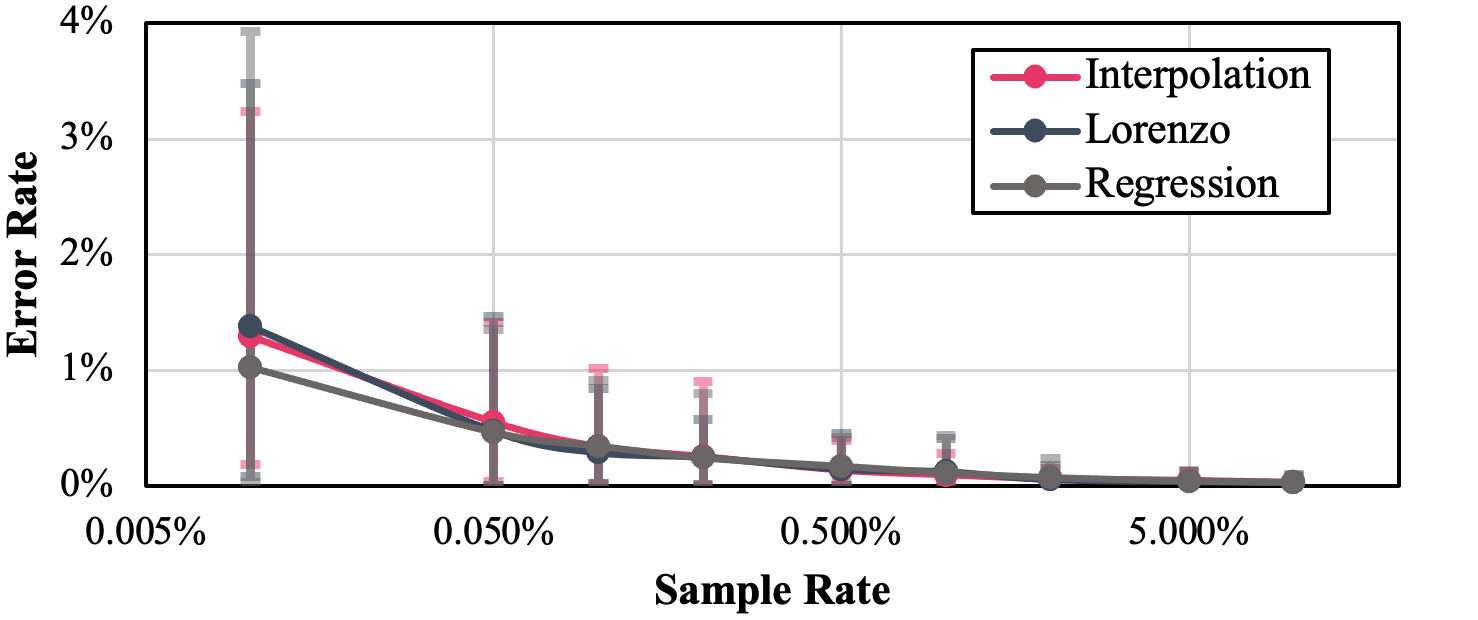}
    \vspace{-1mm}
    \caption{Error rate between sampled prediction error and original prediction error under different sampling rates with three predictors. The error bar indicates the max and min values.}
    \label{fig:fig-eval-sample}
    % \vspace{-4mm}
\end{figure}

To evaluate the effectiveness of our sampled prediction error, we compare the standard deviation of the sampled data and the overall data. Figure~\ref{fig:fig-eval-sample} shows the decreasing error rate with increasing sampling rate with multiple predictors. We can observe that different predictors behave similarly in terms of error rate  with the same sampling rate. 
The sampled prediction error is used for compression ratio modeling and requires high fidelity to the original to provide accurate estimation. 
Based on our experiment, we choose the sampling rate of 1\% in this paper to balance between the sampled data accuracy and sampling overhead.
The detailed sampling error on all datasets with 1\% sampling rate can be found in Table~\ref{tab:Result}. Overall, our sampling strategy can achieve the sample error (i.e., the standard deviation relative to the value range) of only 0.12\% on average.

\subsubsection{Huffman Encoding Efficiency}

\begin{figure}[]
    \centering
    \includegraphics[width=\linewidth]{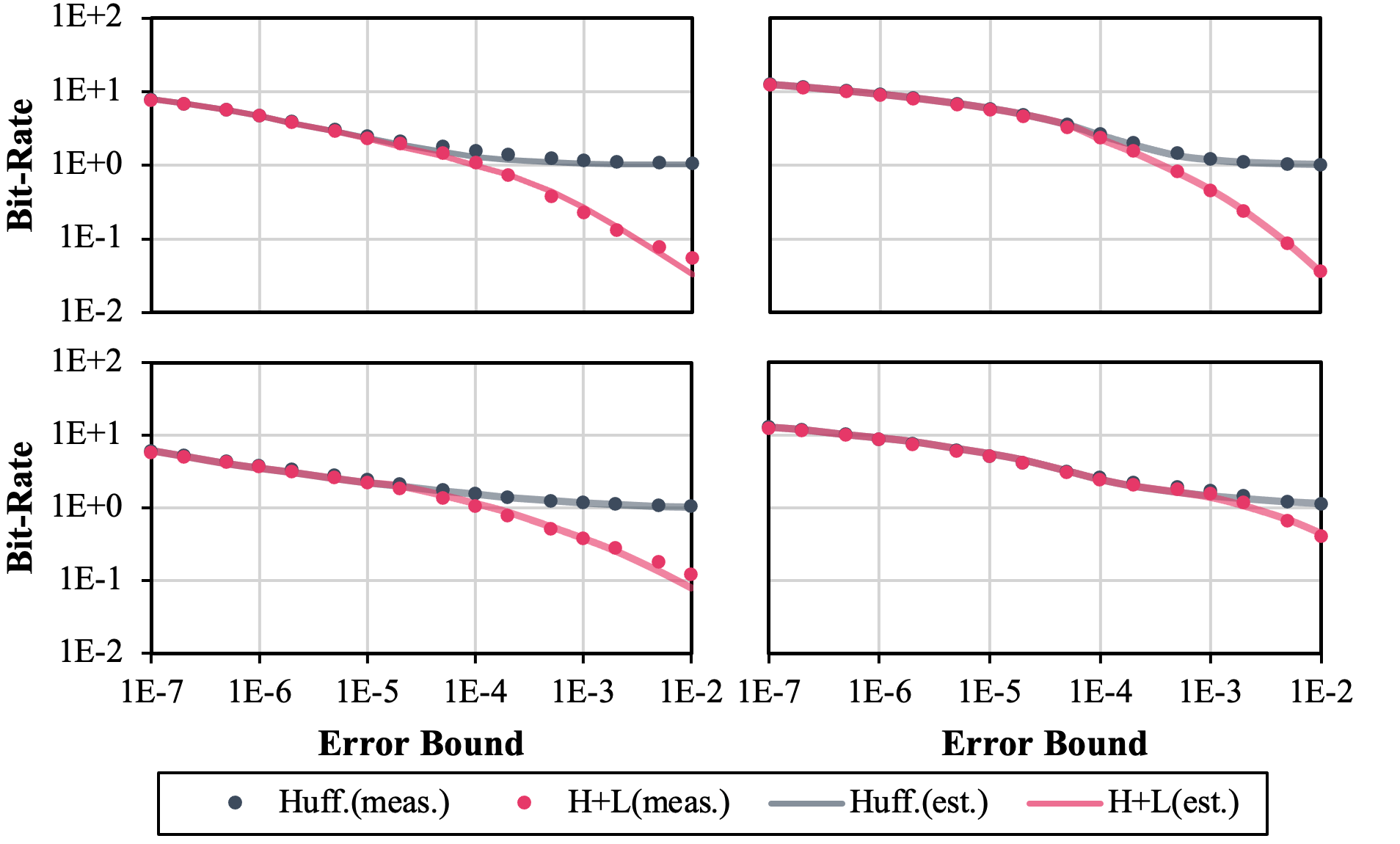}
    % \vspace{-1mm}
    \caption{Compression ratio (bit-rate) estimation accuracy compared to measurement by the encoders.}
    \label{fig:fig-eval-rate}
    % \vspace{-1mm}
\end{figure}

\begin{table*}[]
% \vspace{3mm}
\renewcommand*{\arraystretch}{1.4}
\centering
\ttfamily
% \scriptsize
\footnotesize
\caption{Details of Evaluation Results on Tested Data and Fields}
%\vspace{-4mm}
\newcommand\alignmiddle[2]{
\makebox[3em][r]{$(#1$}\makebox[.8em]{$,\ $}\makebox[2em][l]{$#2)$}}

\newcommand{\BARrev}[1]{%
\begin{tikzpicture}
    \draw[draw=none, fill=black!80] (0,0) rectangle ++({-#1*.12}, -0.04);
    \node[anchor=south east, inner sep=0pt, yshift=.1ex] {#1\%};
\end{tikzpicture}%
}
\newcommand{\BARels}[1]{%
\begin{tikzpicture}
    \draw[draw=none, fill=black!00] (0,0) rectangle ++(0.03, -0.04);
    \node[anchor=south east, inner sep=0pt, yshift=.1ex] {#1};
\end{tikzpicture}%
}

\newcommand{\SUPERBOLD}{\fontfamily{ugq}\selectfont}

\resizebox{0.90\linewidth}{!}{
\begin{tabular}{l|cc|rrrrrr}
\hline 
\textbf{Name}                       & \textbf{Field}       & \textbf{Dim}           & \textbf{Sample Err.} & \textbf{Huff Err.} & \textbf{Lossless Err.} & \textbf{Huff+LL. Err.} & \textbf{PSNR Err.} & \textbf{SSIM Err.} \\ \hline 
\rule{0pt}{1.5\normalbaselineskip}
\multirow{3}{*}{\hspace{-1ex}\BARels{RTM}}       & \BARels{1000}        & \BARels{235x449x449}   & \BARels{0.03\%}       & \BARrev{5.67}     & \SUPERBOLD\BARrev{9.82}    & \BARrev{8.72}         & \BARrev{0.77}     & \SUPERBOLD\BARrev{9.34}     \\
                                    & \BARels{2000}        & \BARels{235x449x449}   & \BARels{0.02\%}       & \BARrev{3.32}     & \SUPERBOLD\BARrev{9.01}    & \BARrev{7.76}         & \BARrev{1.56}     & \SUPERBOLD\BARrev{6.56}     \\
                                    & \BARels{3000}        & \BARels{235x449x449}   & \BARels{0.06\%}       & \BARrev{1.88}     & \SUPERBOLD\BARrev{9.15}    & \BARrev{7.57}         & \BARrev{2.84}     & \SUPERBOLD\BARrev{4.12}     \\
\multirow{2}{*}{\BARels{CESM}}      & \BARels{TS}          & \BARels{1800x3600}     & \BARels{0.06\%}       & \BARrev{6.88}     & \SUPERBOLD\BARrev{11.26}   & \BARrev{8.85}         & \SUPERBOLD\BARrev{3.97}     & \BARrev{2.54}     \\
                                    & \BARels{TROP\_Z}     & \BARels{1800x3600}     & \BARels{0.20\%}       & \BARrev{7.56}     & \SUPERBOLD\BARrev{10.52}   & \BARrev{9.66}         & \BARrev{2.97}     & \SUPERBOLD\BARrev{4.44}     \\
\multirow{2}{*}{\BARels{Hurricane}} & \BARels{U}           & \BARels{100x500x500}   & \BARels{0.10\%}       & \SUPERBOLD\BARrev{4.62}     & \BARrev{3.46}    & \BARrev{5.75}         & \BARrev{1.56}     & \SUPERBOLD\BARrev{5.43}     \\
                                    & \BARels{TC}          & \BARels{100x500x500}   & \BARels{0.12\%}       & \SUPERBOLD\BARrev{5.44}     & \BARrev{2.96}    & \BARrev{5.95}         & \BARrev{2.42}     & \SUPERBOLD\BARrev{3.80}     \\
\multirow{3}{*}{\BARels{Nyx}}       & \BARels{Dark Matter} & \BARels{512x512x512}   & \BARels{0.14\%}       & \SUPERBOLD\BARrev{7.53}     & \BARrev{4.36}    & \BARrev{7.67}         & \BARrev{1.78}     & \SUPERBOLD\BARrev{6.55}     \\
                                    & \BARels{Temperature} & \BARels{512x512x512}   & \BARels{0.13\%}       & \BARrev{3.92}     & \SUPERBOLD\BARrev{5.13}    & \BARrev{3.99}         & \BARrev{1.89}     & \SUPERBOLD\BARrev{4.34}     \\
                                    & \BARels{Velosity Z}  & \BARels{512x512x512}   & \BARels{0.07\%}       & \BARrev{6.85}     & \SUPERBOLD\BARrev{8.65}    & \BARrev{8.08}         & \BARrev{2.64}     & \SUPERBOLD\BARrev{3.90}     \\
\multirow{2}{*}{\BARels{HACC}}      & \BARels{xx}          & \BARels{280953867}     & \BARels{0.26\%}       & \SUPERBOLD\BARrev{2.29}     & \BARrev{1.34}    & \BARrev{3.22}         & \BARrev{1.98}     & \BARels{-}        \\
                                    & \BARels{vx}          & \BARels{280953867}     & \BARels{0.27\%}       & \SUPERBOLD\BARrev{3.71}     & \BARrev{1.49}    & \BARrev{3.83}         & \BARrev{3.67}     & \BARels{-}        \\
\BARels{Brown}                      & \BARels{Pressure}    & \BARels{8388609}       & \BARels{0.11\%}       & \SUPERBOLD\BARrev{5.99}     & \BARrev{5.68}    & \BARrev{6.46}         & \BARrev{4.42}     & \BARels{-}        \\
\BARels{Miranda}                    & \BARels{vx}          & \BARels{256x384x384}   & \BARels{0.13\%}       & \SUPERBOLD\BARrev{7.90}     & \BARrev{6.95}    & \BARrev{8.71}         & \BARrev{2.55}     & \SUPERBOLD\BARrev{8.92}     \\
\BARels{QMCPACK}                    & \BARels{einspine}    & \BARels{69x69x115}     & \BARels{0.13\%}       & \BARrev{6.84}     & \SUPERBOLD\BARrev{8.83}    & \BARrev{6.20}         & \BARrev{5.67}     & \SUPERBOLD\BARrev{7.43}     \\
\BARels{SCALE}                      & \BARels{PRES}        & \BARels{98x1200x1200}  & \BARels{0.16\%}       & \BARrev{1.65}     & \SUPERBOLD\BARrev{2.79}    & \BARrev{2.36}         & \BARrev{1.72}     & \SUPERBOLD\BARrev{5.35}     \\
\BARels{EXAFEL}                     & \BARels{raw}         & \BARels{10x32x185x388} & \BARels{0.12\%}       & \SUPERBOLD\BARrev{5.64}     & \BARrev{4.25}    & \BARrev{6.23}         & \BARrev{3.80}     & \BARels{-}        \\ \hline
\SUPERBOLD Average                   & -           & -             &  0.12\%       &  5.16\%     & \SUPERBOLD 6.21\%    &  6.53\%         &  2.72\%    & \SUPERBOLD 5.59\%     \\ \hline
\end{tabular}
}

\begin{tablenotes}
    %   \scriptsize
    %   \flushright
    \centering
      \item \scriptsize \hspace{14mm} \textcolor{black}{* Bold items highlight the larger prediction error between the two encoders and between the two post analyses}
\end{tablenotes}
\vspace{-5mm}
\label{tab:Result}
\end{table*}

Next, we evaluate our modeling of Huffman coding efficiency. The dark black dot and line in Figure~\ref{fig:fig-eval-rate} shows the measured bit rate after Huffman encoding and the estimated bit rate. The modeling matches the measurements very well above bit-rate of about 2 bits based on Equation (\ref{equ-2}). After this point, the model switches to the fitted function based on the three anchor points. The lowest estimated bit rate threshold is 1 bit, as expected for Huffman coding in extreme situations. 
To quantify the accuracy of our modeled Huffman efficiency, we introduce the error computation based on the standard deviation of the ratio between estimated values and measured values:
\begin{align}
\small\textstyle
    E = 1-(1+STD(\frac{R}{R'}-1))^{-1},
    \label{equ-eval} 
\end{align}
where $E$ is the accuracy, and $R$ and $R'$ are the measured values and estimated values, respectively.
This equation is also used to quantify the prediction error of the following evaluations in this paper. 
% including the lossless efficiency estimation accuracy, overall compression ratio estimation accuracy, post-hoc analysis estimation accuracy. %%%
Based on Equation (\ref{equ-eval}), the detailed Huffman encoding efficiency estimation accuracy on all datasets can be found in Table~\ref{tab:Result}. Overall, our Huffman encoding model exhibits a high accuracy of up to 98.4\% and 94.8\% on average.
\textcolor{black}{Here we illustrate the error rate, while it can be easily converted to the accuracy (e.g., error rate of 5.16\% means the prediction accuracy of 94.84\% for Huffman coding).}

\subsubsection{RLE Efficiency}

% Run-length encoding, as mentioned in Section~\ref{sec:encoder}, can provide extra compression by lossless compression of the Huffman encoder output. 
We compare our model to the extra compression ratio provided by Zstandard lossless compressor.
From Table~\ref{tab:Result}, our model and assumption can accurately provide the compression ratio from the extra lossless compression.
The accuracy is up to 98.5\% and is 93.8\% on average,
\textcolor{black}{which is worse than the prediction with only Huffman coding, due to the approximation from lossless compression to RLE.}

\subsubsection{Overall Compression Ratio}

The overall compression ratio is the combination of Huffman encoding efficiency and RLE efficiency. 
The red dot and line in Figure~\ref{fig:fig-eval-rate} shows the measured bit rate and the estimated bit rate of overall encoder efficiency, respectively. 
% with both Huffman encoder and the optional Zstandard. 
We can observe that our modeling achieves a high accuracy compared to the measurements.
Detailed accuracy of overall compression ratio estimation on all datasets can be found in Table~\ref{tab:Result}. 
% Overall, our model achieves up to 97.6\% and on average 93.5\% of accuracy for modeling the compression ratio.
The accuracy is up to 97.6\% and is 93.5\% on average.
\textcolor{black}{The result shows that our compression-ratio estimation with only Huffman coding is almost always more accurate than that with both Huffman-coding and lossless-encoding stages.
Moreover, we observe that our model performs slightly differently across datasets. For example, the estimation errors of Huffman coding and lossless encoding on the HACC dataset are lower than on the other datasets. This is because HACC is 1D data and its quantization codes are more randomly distributed, which lowers the possibility of false quantization code prediction caused by using the original values, thus the compression ratio of HACC is easier to predict.
}
\textcolor{black}{Compared to the previous estimation approach~\cite{wang2019compression} with the accuracy of about 40\%$\sim$90\%, our theoretical approach provides much higher estimation accuracy consistently across different datasets.}
% Note that we eliminated the results that contains compression ratio of over $500\times$ because their bit-rate is extremely small and are rarely needed by users. %%%

\subsection{Accuracy of Post-Hoc Analysis Quality Model}

In this subsection, we evaluate the accuracy of our modeling of post-hoc analysis, including on PSNR, SSIM and data specific post-hoc analysis such as FFT.

\subsubsection{PSNR}

\begin{figure}[]
    \centering
    \includegraphics[width=0.98\linewidth]{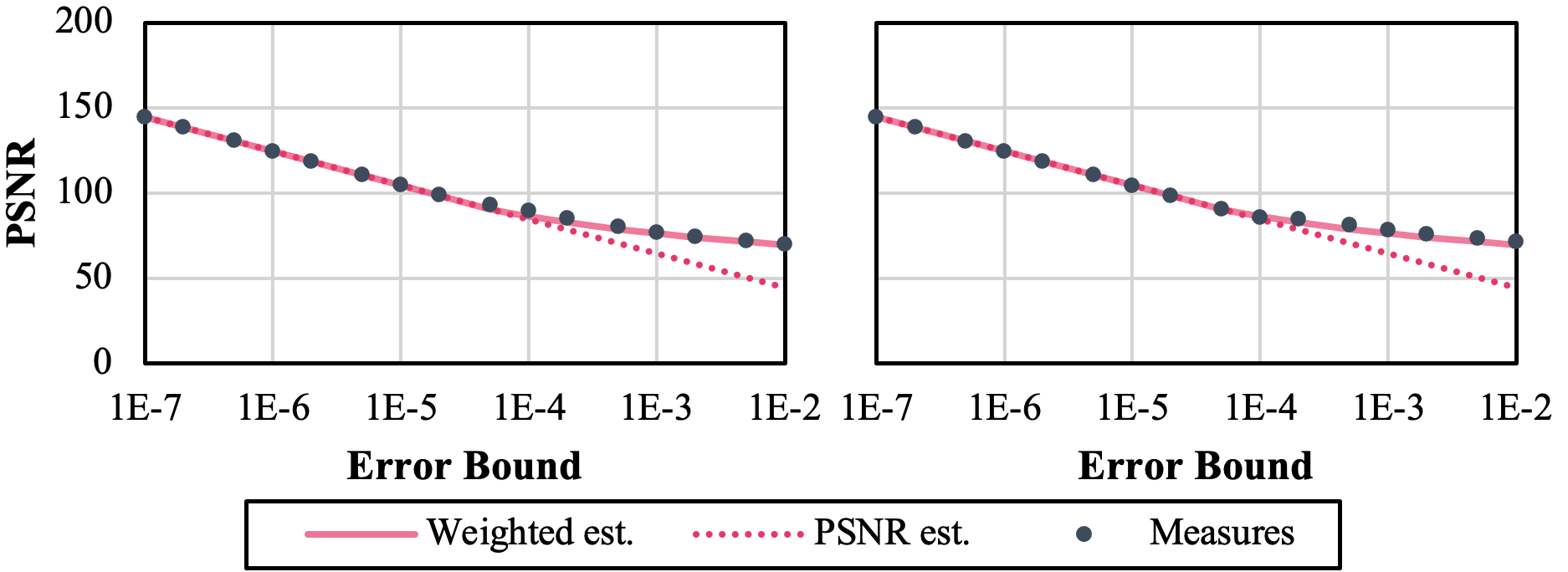}
    % \vspace{-1mm}
    \caption{PSNR estimation accuracy compared to measurement. Evaluated on Nyx dark matter density field with both Linear Interpolation predictor (left) and Lorenzo predictor (right).}
    \label{fig:fig-eval-PSNR}
    % \vspace{-1mm}
\end{figure}

Figure~\ref{fig:fig-eval-PSNR} shows the measured PSNR compared to the estimated PSNR based on the error distribution. 
The dashed red line is the PSNR estimation based on the error distribution defined by Equation (\ref{equ-error-1}), which only considers the uniform distribution.
The solid red line is the PSNR estimation that utilizes both Equations (\ref{equ-error-1}) and (\ref{equ-error-2}).
We can observe that under high error bound situations, the refined distribution of Equation (\ref{equ-error-2}) can benefit the refinement of post-hoc analysis quality estimation. Similar observation can also be found for SSIM and FFT analysis.
Similarly, the detailed PSNR estimation accuracy for all datasets can be found in Table~\ref{tab:Result}.
Overall, our model achieves up to 99.2\% and on average 97.3\% of accuracy for modeling the PSNR.

\subsubsection{SSIM}

\begin{figure}[]
    \centering
    \vspace{-4mm}
    \includegraphics[width=\linewidth]{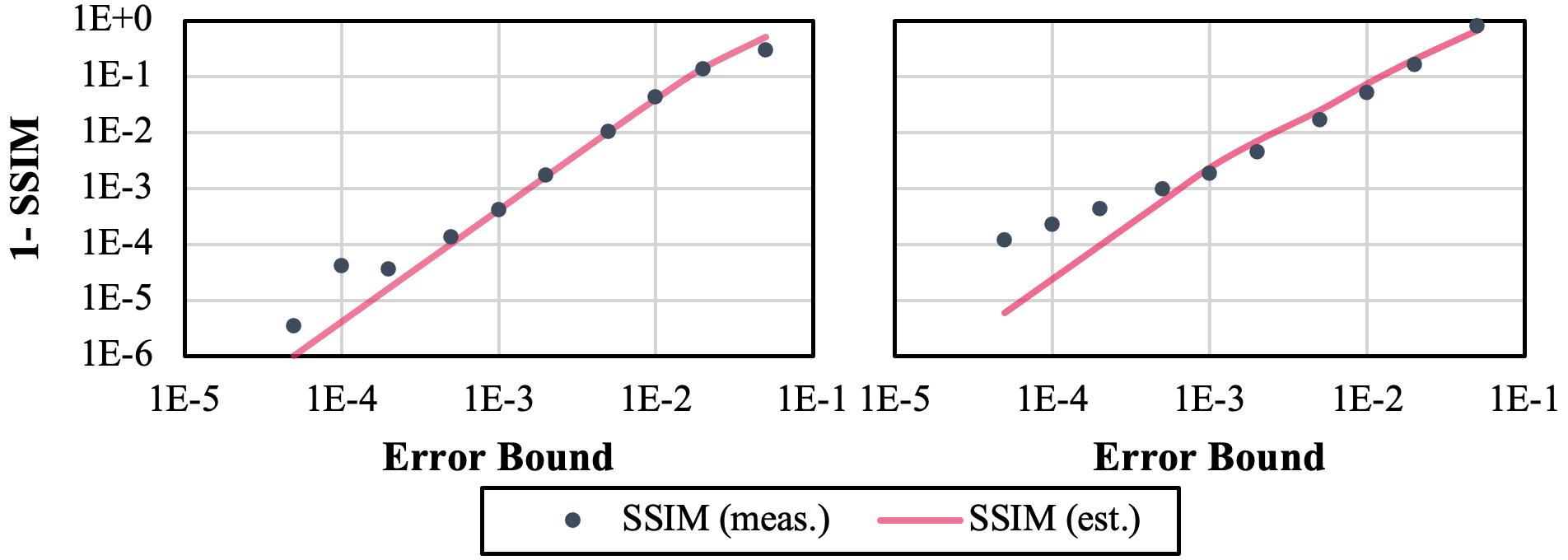}
    % \vspace{-1mm}
    \caption{SSIM estimation accuracy compared to measurement. Evaluated on CESM dataset (left) and Aramco RTM dataset (right).}
    \label{fig:fig-eval-SSIM}
    % \vspace{-4mm}
\end{figure}

Figure~\ref{fig:fig-eval-SSIM} shows the measured SSIM compared to the estimated SSIM based on the error distribution. 
Note we use the ($1-SSIM$) in log scale on y-axis to show the difference between estimation and measurement under lower error bound.
% The parameter $C_2$ in Equation (\ref{equ-ssim-4}) is also adjusted based on each dataset to better demonstrate the value. \textcolor{red}{[Franck]: to better what? I don't understand} %%%
% Sian: Yes, this sentence is confusing, I removed it. C_2 is a constant parameter that can be adjusted in SSIM computation, we adjust it so that the SSIM value is not always close to 1 (SSIM=1 means no difference) and undistinguishable.
The estimation is slightly off under lower error bounds. This is because the approximation we used for Equation (\ref{equ-ssim-2}) is less accurate 
% since the deviations' ratio is less significant compared to the close-to-one but existing averages' ratio.
since the terms we simplified in Equation~(\ref{equ-ssim-3}) \& (\ref{equ-ssim-4}) are no longer negligible in the case where Equation~(\ref{equ-ssim-2}) is close-to-one.
On the other hand, under very high error bounds, the estimation is also degraded since the approximation in Equations (\ref{equ-ssim-3}) and (\ref{equ-ssim-4}) are less accurate when $E[i]$ is larger.
Detailed SSIM estimation accuracy for all datasets can be found in Table~\ref{tab:Result}.
Our evaluation shows that our model on SSIM can provide an accuracy of 94.4\% on average.
\textcolor{black}{Compared to the SSIM estimation, our model performs better on the PSNR estimation.}

\subsubsection{Data Specific Post-hoc Analysis}

\begin{figure}[]
    \centering
    \includegraphics[width=0.93\linewidth]{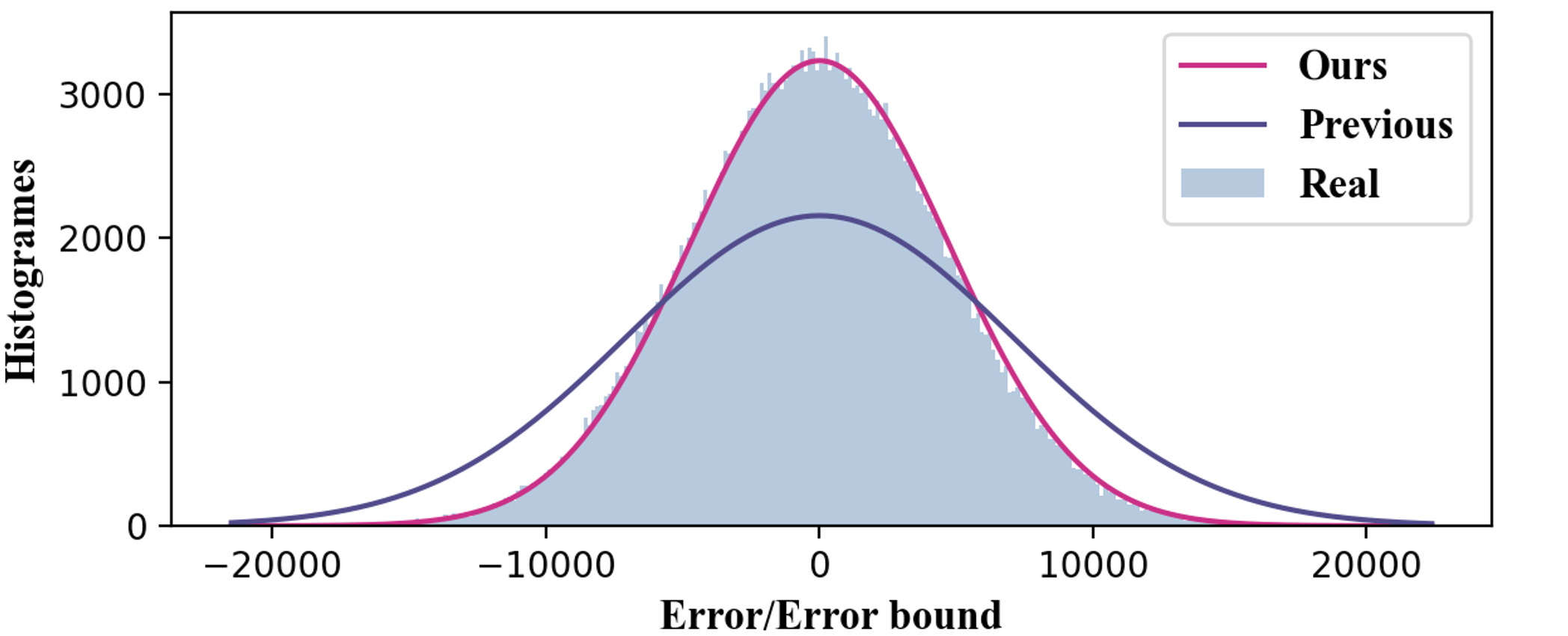}
    % \vspace{-1mm}
    \caption{FFT quality degradation estimation compared to measurement. Evaluated on Nyx temperature field at ABS 500.}
    \label{fig:fig-eval-fft}
    \vspace{-4mm}
\end{figure}

Previous study shows cosmology specific post-hoc analysis quality modeling with the SZ lossy compressor~\cite{jin2020adaptive}. However, it only considered uniform error distribution. 
With our proposed modeling and guideline for post-hoc analysis quality modeling, we can also accurately estimate the FFT quality degradation under high error bound situations. Figure~\ref{fig:fig-eval-fft} shows that the proposed estimation that considers error distribution from both Equations (\ref{equ-error-1}) and (\ref{equ-error-2}) outperforms previous solution that only considered uniform error distributions.

\subsection{Evaluation on Performance Overhead of Our Modeling}

\begin{figure}[]
    \centering
    \includegraphics[width=0.93\linewidth]{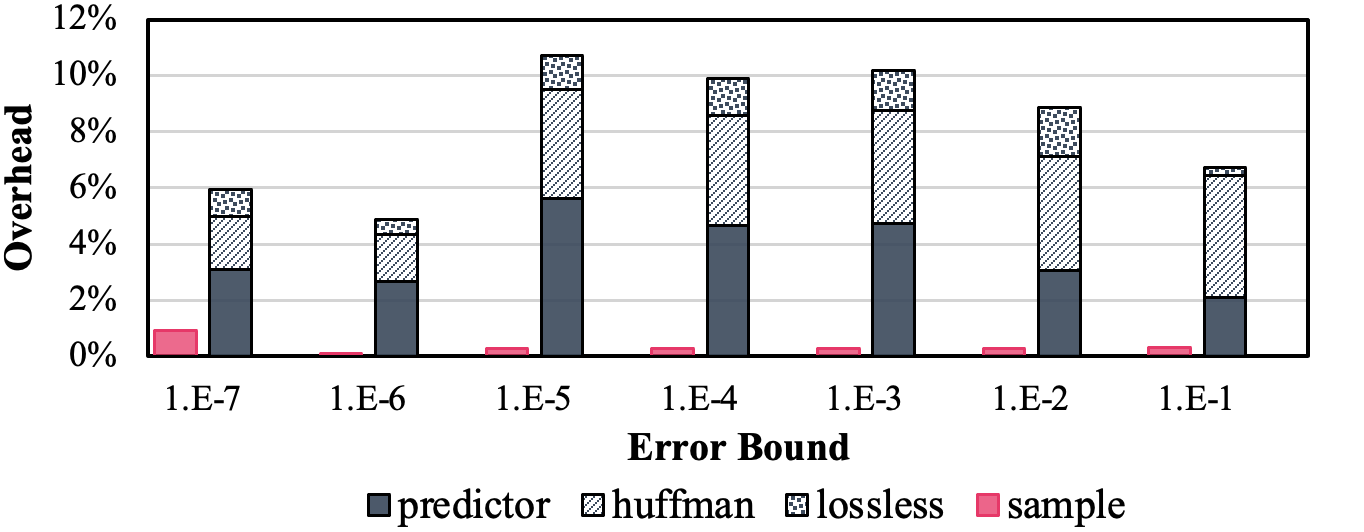}
%    \vspace{-4mm}
    \caption{Performance comparison between proposed modeling solution and previous trial-and-error approach.}
    \label{fig:fig-eval-perf}
    % \vspace{-4mm}
\end{figure}

We compared the performance of our modeling strategy with the trial-and-error approach. In SZ3, to optimize the predictor for a given error bound, the lossy compressor sampled a proportion of data blocks in given dataset and pre-compresses the structured sampling data with multiple predictor candidates. This module can also be used to just estimate the compression ratio with specified error bounds.
When conducting the common use-cases of evaluating the error bound to ratio of a given dataset, our modeling only requires one time data sampling and computes the estimation.
However, the previous trial-and-error must re-compress for \textcolor{black}{each combination of error bound and predictor}.
Figure~\ref{fig:fig-eval-perf} shows the performance comparison between our workflow and the previous approach on average across 3 Reverse Time Migration (RTM) datasets.
Our solution outperforms the trial-and-error solution by 18.7$\times$ on average when considering 7 candidate error bounds to estimate \textcolor{black}{with the Lorenzo and interpolation predictors as candidates}.
\textcolor{black}{Note that the overhead is relative to the overall compression time. %with selected error bound and predictor.
}
We can observe that by running the compression process, the trial-and-error solution spends a large amount of time on Huffman encoding and lossless compression.
Moreover, our solution spends less time even compared to only the predictor part of previous solution, thanks to our newly designed sampling strategy that allows lower sample rates with even higher accuracy.
\textcolor{black}{Note that for each predictor our sampling only happens once (at the error bound of 1E-7) for all error bounds, since we can compute the distribution of quantization codes based on our model instead of repeating the prediction with different error bounds.
It is worth noting that the trial-and-error approach cannot utilize our new sampling strategy due to the difference of using the original values and the reconstructed values in prediction.}
In addition, the overhead of our solution is relatively stable across all tested datasets, which demonstrates our consistent high performance in parallel applications (with multiple processes/datasets) using lossy compression.

\subsection{Use-Cases Study}

In this section, we investigate the effectiveness of utilizing our ratio-quality model for the three use-cases on the 3D RTM dataset \cite{rtm}.
RTM is an important seismic imaging method for oil and gas exploration~\cite{zhou2018reverse,alturkestani2020maximizing}.
It has forward modeling and backward propagation stages that write and then read the state of the computed solution at specific timesteps, so lossy compression is used to significantly reduce the I/O and memory overheads for 3D RTM.
%the overhead in migration tasks.
%As mentioned in Section~\ref{sec:design}, we demonstrate our ratio-quality model on three use cases. %%%
%in Section~\ref{sec:design}. 

\subsubsection{Predictor Selection}

\begin{figure}[]
    \centering
    \includegraphics[width=0.92\linewidth]{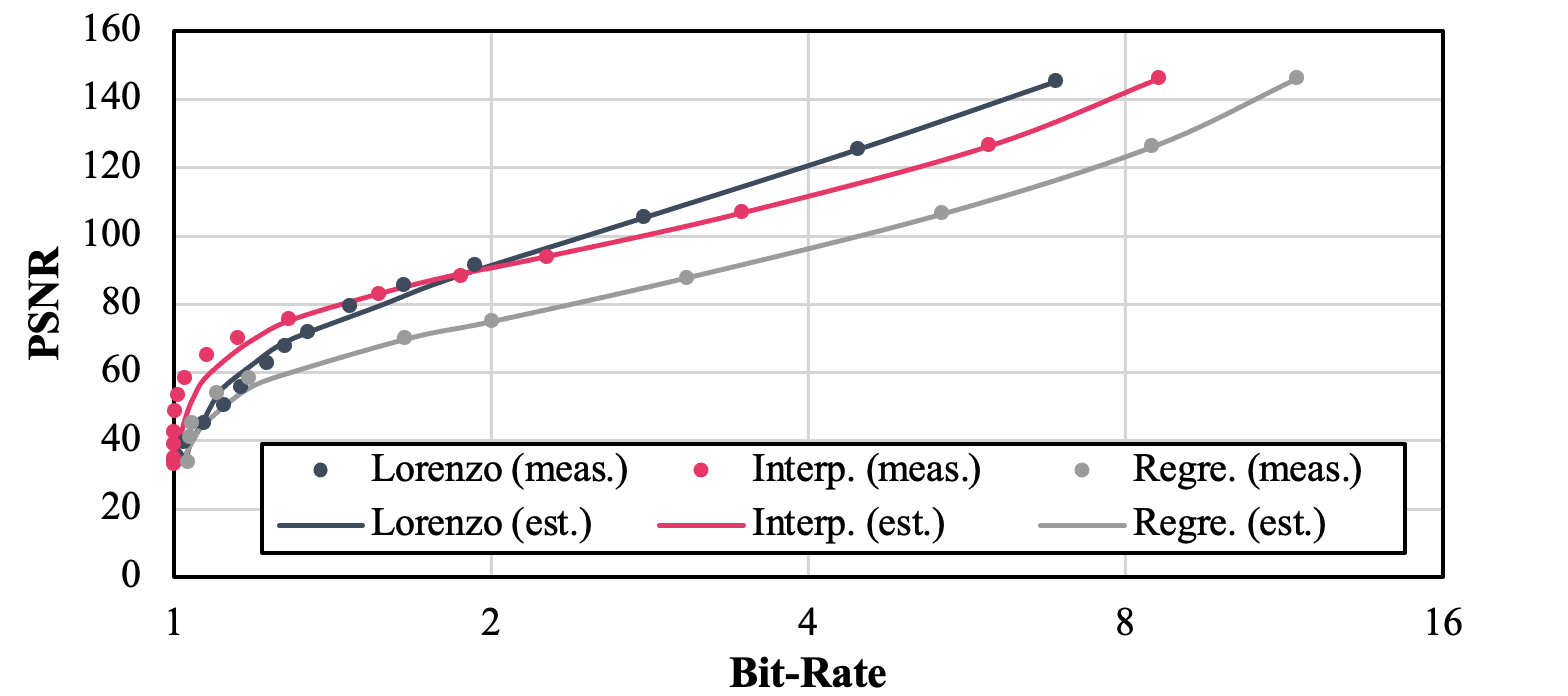}
    \vspace{-1mm}
    \caption{Rate-distortion curve of multiple predictors with different error bound. Evaluated with RTM dataset.}
    \label{fig:fig-eval-predictor}
    % \vspace{-4mm}
\end{figure}

For the first use-case, we perform our ratio-quality model on RTM 3D dataset with all three predictors and use PSNR as the analysis metric.
Figure~\ref{fig:fig-eval-predictor} shows the estimation and measured rate-distortion correlation. First, we can observe that the estimated rate-distortion curve based on our model is highly accurate compared to measured data points. Secondly, we can clearly see that the linear interpolation predictor provides higher PSNR with the same bit-rates under lower bit-rate ranges. Our model estimation suggests to switch to the linear interpolation from the Lorenzo predictor as the preferred predictor when the estimated bit-rate is lower than 1.89. This estimation is accurate compared to the measured bit-rates for the predictor switch between [1.47, 1.93]. 
Our solution also provides 21.8$\times$ performance improvement by reducing the overhead from 109.97\% to 5.04\% compared to the previous sampling solution that requires a trial-and-error for each error bound.

\subsubsection{Memory Limitation Control}

\begin{figure}[]
    \centering
    \includegraphics[width=0.95\linewidth]{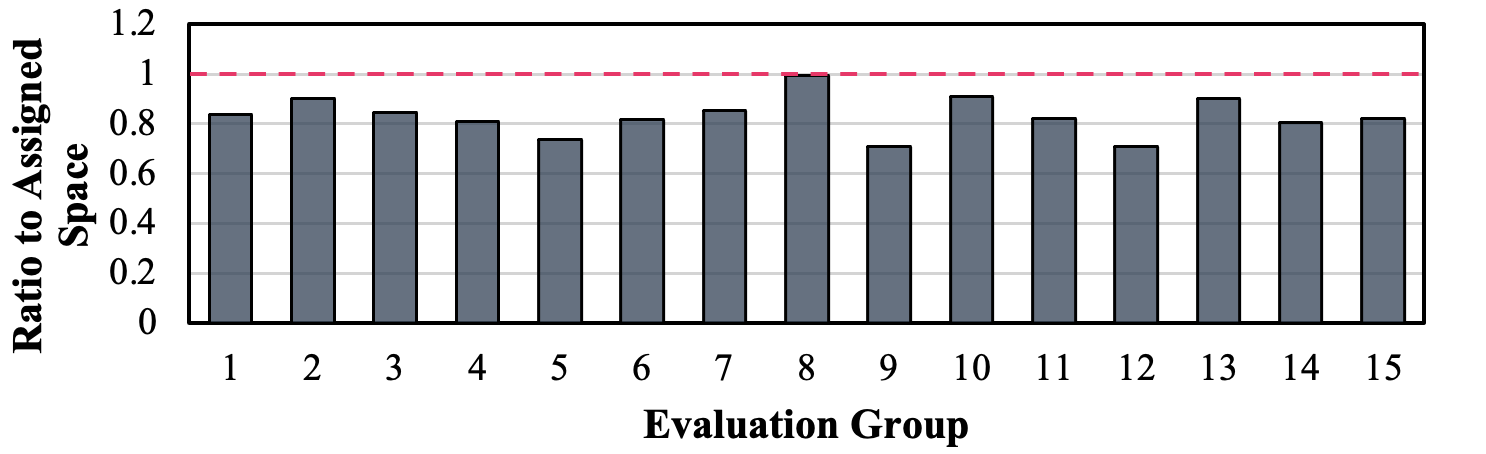}
%    \vspace{-7mm}
    \caption{Ratio of measured space consumption to assigned space. Evaluated with RTM dataset, randomly choose time steps and error bound for 15 groups.}
    \label{fig:fig-eval-memory}
    \vspace{-3mm}
\end{figure}

%%% Since our compression ratio model provides ratio estimation instead of ratio boundary. We cannot select the error-bound and grantee the compressed file smaller than assigned memory space. 
% Thus, our strategy is to downgrade the target bit-rate of our model to 80\% of that satisfies the memory control to allow some uncertainty between estimation and measures. 
Figure~\ref{fig:fig-eval-memory} shows the result of compressed file size relative to the assigned memory based on our compression ratio model. We can observe that although many evaluated groups result in larger file sizes than estimated (i.e., 80\%), they still stay within the assigned space thanks to the high accuracy of our compression ratio model.
The compressed file that exceeds the space limitation may require re-optimization based on a lower target bit-rate.
Considering the low possibility of such a situation (e.g.., around 5\% for RTM dataset) and the low overhead to recompute the error bound with a given target bit-rate, our solution is still highly practical for fix-rate compression.

\subsubsection{In-Situ Compression Optimization}

\begin{figure}[t]
    % \centering
    \includegraphics[width=0.92\linewidth]{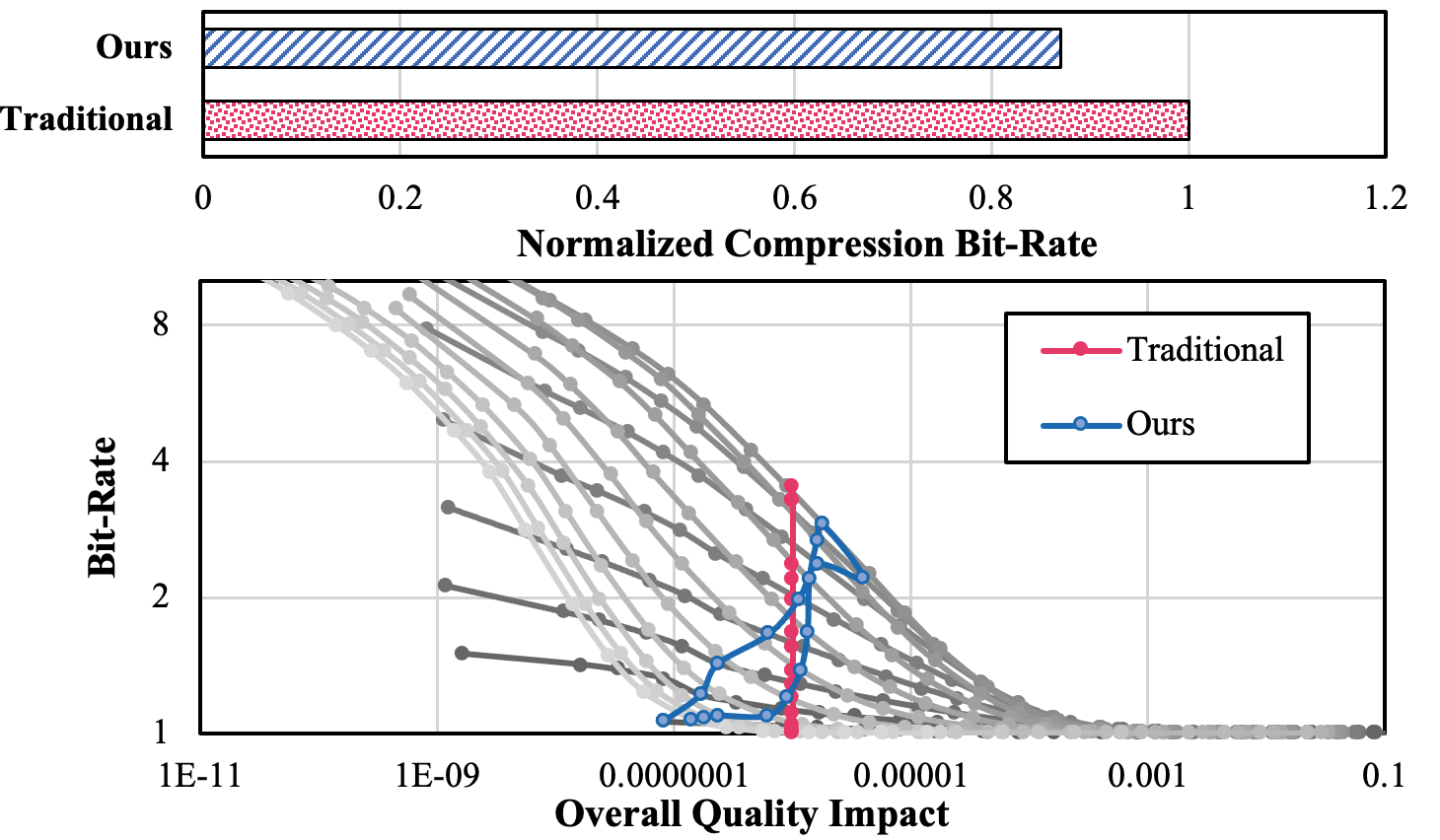}
    % \vspace{-2mm}
    \caption{Error bound optimization for RTM dataset with multiple time steps in consideration for post-hoc analysis.}
    \label{fig:fig-eval-insitu}
    % \vspace{-3mm}
\end{figure}

% We consider the RTM data demonstrating our third use-case. 
The RTM data used for PSNR analysis is a stacked image built form images of multiple timesteps. We consider the images from different timesteps as partitions that form the final dataset. 
% The proposed optimization strategy is used to balance the ratio-quality of data for each partition. %%%
% One of the specific analysis approach for RTM data is to perform the analysis on the sum of the data from multiple timesteps. 
 With our ratio-quality model, we optimize the error bound for each image by balancing the ratio of bit-rate and impact on the overall quality degradation. 
Figure~\ref{fig:fig-eval-insitu} shows the optimized error bounds from our model for every timestep. The quality of each timestep here influences the overall quality of the stacked image. We can observe the trade-offs between timesteps for ratio and quality. Overall, we can provide an extra 13\% of compression ratio with same post-hoc analysis quality, or an extra 31\% of post-hoc analysis quality with the same compression ratio, compared to using the same error bound for all timesteps.
By leveraging our ratio-quality model, we can perform in-situ optimization for parallel applications, which is infeasible with previous solutions.

\subsection{Evaluation on Overall Performance of Data Management} 

\begin{figure}[]
    \centering
    \includegraphics[width=0.92\linewidth]{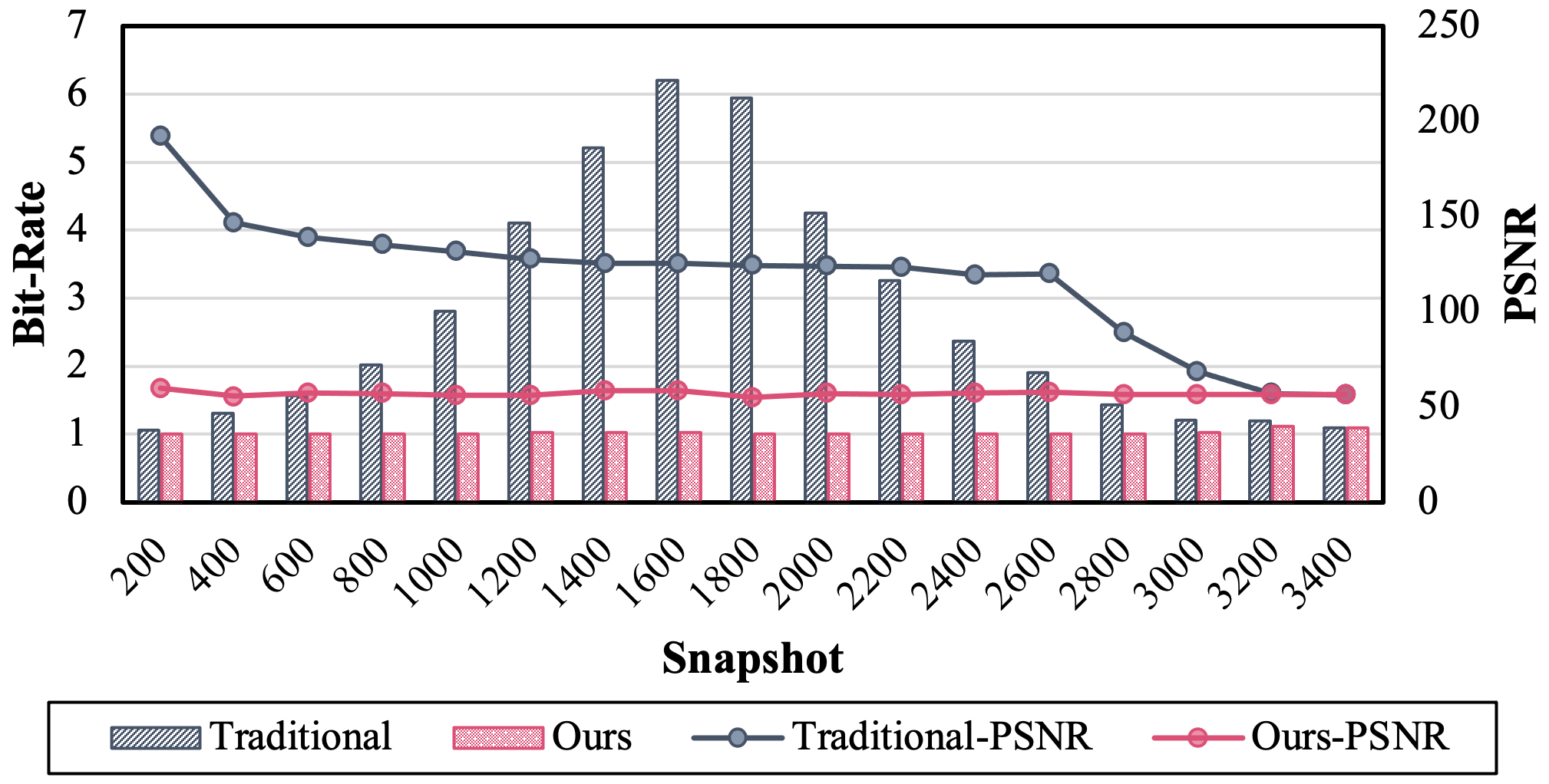}
    % \vspace{-2mm}
    \caption{Comparison between our modeling based method with offline optimization method in terms of both bit-rate and corresponding PSNR across different snapshots when target PSNR is 56 dB.}
    \label{fig:fig-eval-ratio}
    % \vspace{-1mm}
\end{figure}

\textcolor{black}{
In this section, we use the RTM data to demonstrate the effectiveness of our ratio-quality model when deploying it to the scientific data management systems. 
More specifically, we use parallel HDF5 \cite{Parallel8:online} built upon MPI-IO \cite{mpiio/10.1145/301816.301826} as our data management system. 
Similar to other scientific applications, such as Nyx~\cite{nyx} and HACC\cite{hacc}, the RTM simulation needs to store one snapshot every few iterations for
%back propagation and 
future use and analysis. The overall simulation time spent in I/O can easily reach beyond 50\%, thus, lossy compression of data before storing it can significantly improve the I/O performance.
}

\begin{figure*}[]
    \centering
    \includegraphics[width=0.92\linewidth]{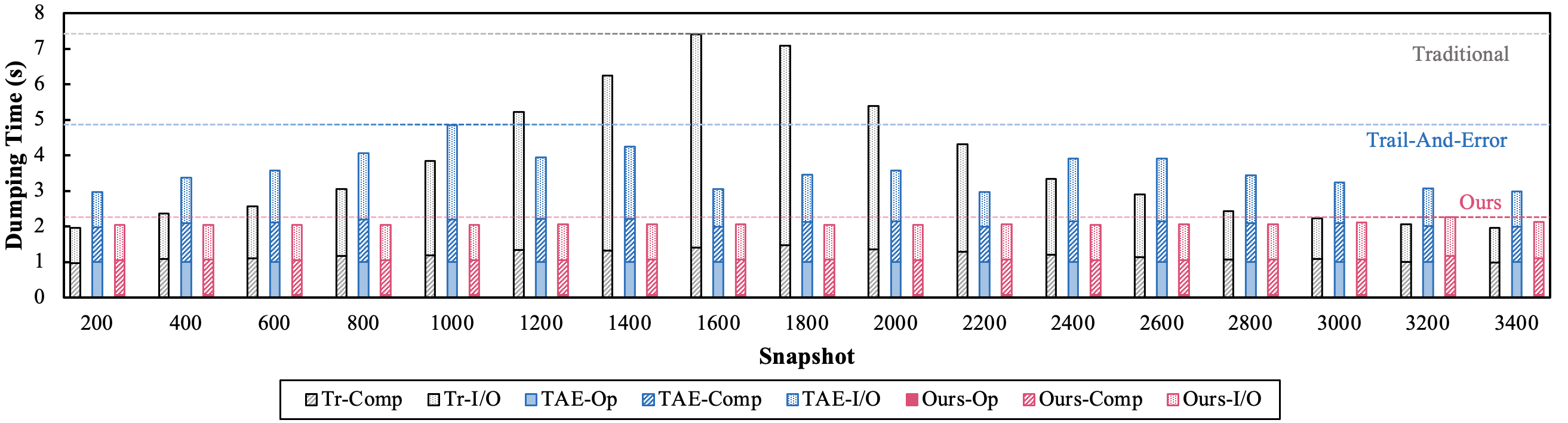}
    % \vspace{-2mm}
    \caption{Overall data dumping performance with parallel HDF5. Comparison between traditional method, trial-and-error and our modeling based method. Dashed lines highlight the maximum dumping time occurred in the simulation. ``Tr'' refers to the traditional approach, ``TAE'' refers to the in-situ trial-and-error approach. `Comp', `I/O', and `Op' refer to times of compression, I/O, and optimization, respectively.}
    \label{fig:fig-eval-overall}
    \vspace{-4mm}
\end{figure*}

\textcolor{black}{
Previously, researchers must conduct offline trial-and-error experiments or use a benchmark toolkit (e.g., Foresight~\cite{foresight-git}) to determine the bestfit compression configurations for a given set of data. 
Such static analysis takes a long time, but to make matters worse, it can only choose the worst case configurations to guarantee all the reconstructed data quality, similar to the Liebig's barrel.
In the following experiments, we call this static offline analysis method the traditional approach.
\textcolor{black}{More specifically, we let the compressor to experiment the error bound from 5 candidates (i.e., ABS 1E-4, 1E-5, 1E-6, 1E-7, 1E-8) for all snapshots and choose one error bound that fits all. During the performance evaluation, we apply this error bound (i.e., ABS 1E-7 based on our experiment) to all snapshots.}
Different from the traditional approach, our ratio-quality model allows us to in-situ determine the optimized error bound for different snapshots based on the desired reconstructed data quality (i.e., PSNR).
Figure~\ref{fig:fig-eval-ratio} shows the ratio-quality comparison on the RTM data across multiple snapshots between the traditional static, offline approach and our adaptive, in-situ approach. 
In this example, our target is to ensure the PSNR is higher than 56 dB for all snapshots, which guarantees the reconstructed data quality of every snapshot for postprocessing stages.
We can observe that the traditional solution chooses only one error bound for all snapshots, causing the PSNR of most snapshots to be much higher than the target. 
%ICDE_revision% This wastes the potential to further compress the data, but only to ensure the reconstruction quality of the snapshots from 3200 to 3400.
By comparison, our in-situ solution with the ratio-quality model can provide consistent and low bit-rate across all snapshots while satisfying the requirement on reconstruction data quality.
}

\textcolor{black}{
Other than the traditional approach, we also implement the trial-and-error method as an in-situ process in our data management system for a fair comparison, referred to as the in-situ trial-and-error (TAE) approach in the following experiment.
More specifically, we let the compressor to experiment the error bound from 5 candidates 
% ICDE_revision% \sout{(i.e., ABS 1E-4, 1E-5, 1E-6, 1E-7, 1E-8)}
for a given snapshot before actually compressing it.
\textcolor{black}{Different from the traditional offline approach, we preform this optimization online and choose the optimal error bound for a given snapshot, which introduces an additional optimization overhead.}
Figure~\ref{fig:fig-eval-overall} shows the overall data dumping time of all three methods across different snapshots: the traditional approach, the in-situ TAE approach, and our in-situ optimized approach with the ratio-quality model.
This experiment is conducted with 128 processes on 8 nodes, each process holding a portion of each snapshot.
It includes three types of time: (1) optimization time, the time spent on compression configuration optimization, including the experimenting time in the in-situ TAE approach; (2) compression time, the time spent for compressing the data with specified configuration; and (3) I/O time, the time spent to actually store the compressed data with parallel HDF5.
Note that the baseline data-dumping time without any compression is 29.4s \textcolor{black}{for each snapshot}, which is higher than any of the three approaches.
Compared with the traditional and the in-situ TAE approaches, our approach can significantly reduce the overall data-dumping time thanks to the accurate error bound control that provides high compression ratio.
When specifically compared with the in-situ TAE approach, our approach can significantly reduce the optimization time while providing the higher compression ratio to reduce the I/O time.
This is because the in-situ TAE approach must experiment several configurations, that is not only time consuming, but also provides limited error bound granularity.
Moreover, note that the dumping time of our solution is highly stable, and the longest dumping time is noticeably lower than that of the other two methods, which is critical for the overall stable throughput of the data management system.
%potentially helpful to overlapping.
Overall, our optimization solution with the proposed ratio-quality model can reduce the data management time by up to 3.4$\times$ compared to the traditional static, offline solution and by up to 2.2$\times$ compared to the in-situ trial-and-error implementation.
} 
\section{Conclusion and Future Work}
\label{sec:conclusion}

In this paper, we develop a general-purpose analytical ratio-quality model for prediction-based lossy compressors that can effectively estimate the compression ratio, as well as the impact of the lossy compressed data on post-hoc analysis quality.
Our analytical model significantly improves the prediction-based lossy compression in three use-cases:
(1) optimization of predictor and compression mode by selecting the best-fit predictor and mode automatically; 
(2) memory compression optimization by selecting error bounds for fixed or estimated bit-rates; 
and (3) overall ratio-quality optimization by fine-grained error-bound tuning of various data partitions. 
We evaluate our analytical model on 10 scientific datasets, demonstrating its high accuracy (93.47\% accuracy on average) and low computational cost (up to 18.7$\times$ lower than the previous approach) for estimating the compression ratio and the impact of lossy compression on post-hoc analysis quality. 
We also verify high effectiveness of our ratio-quality model using different applications across the three use-cases.
Finally, we demonstrate that our modeling based approach reduces the data management time for the RTM simulation by up to 3.4$\times$ with parallel HDF5 on 128 CPU cores, compared to the traditional static, offline solution. 
\textcolor{black}{In the future, we plan to extend our model to other lossy compressors such as the transform-based lossy compressor ZFP~\cite{zfp} and more post-hoc analysis metrics. In addition, we also plan to target more aggressive memory control with higher memory usage compared to 80\% used in the current approach.}

\section*{Acknowledgments}

\small This research was supported by the Exascale Computing Project (ECP), Project Number: 17-SC-20-SC, a collaborative effort of two DOE organizations---the Office of Science and the National Nuclear Security Administration, responsible for the planning and preparation of a capable exascale ecosystem, including software, applications, hardware, advanced system engineering and early testbed platforms, to support the nation's exascale computing imperative. The material was supported by the U.S. Department of Energy, Office of Science, Advanced Scientific Computing Research (ASCR), under contracts DE-AC02-06CH11357 and DE-AC02-05CH11231. This work was also supported by the National Science Foundation under Grants OAC-2003709, OAC-2042084, OAC-2104023, and OAC-2104024. We gratefully acknowledge the computing resources provided %on Bebop, a high-performance computing cluster operated 
by the Argonne Laboratory Computing Resource Center.

%\newpage
% \bibliographystyle{ACM-Reference-Format}
\bibliographystyle{IEEEtran}
\bibliography{refs}

\end{document}